\documentclass[twocolumn,tighten,twocolappendix,emulateapj]{aastex701}
\usepackage{natbib}
\usepackage{color,ulem}
\usepackage{xcolor}
\usepackage{graphicx} %
\usepackage[caption=false]{subfig}
\usepackage[T1]{fontenc}

\newcommand{\dd}{\mathrm{d}}

\newcommand{\shellavg}[1]{\langle #1 \rangle_{\theta, \phi}}
\newcommand{\shellint}[1]{\left\{ #1 \right\}_{\theta, \phi}}

\newcommand{\polaravg}[1]{\langle #1 \rangle_{\theta}}
\newcommand{\polarint}[1]{\left\{ #1 \right\}_{\theta}}

\newcommand{\timeazimavg}[1]{\langle #1 \rangle_{t, \phi}}
\newcommand{\azimavg}[1]{\langle #1 \rangle_{\phi}}

\newcommand{\timeavg}[1]{\langle #1 \rangle_{t}}

\usepackage{amsmath}	%
\usepackage{amssymb}	%
\usepackage{cases}
\usepackage{array,multirow}
\usepackage{upgreek}
\usepackage{xspace}
\usepackage{mathtools}
\usepackage{siunitx}
\usepackage[skip=0pt]{caption}
\usepackage{hyperref}
\usepackage{placeins}

\begin{document}

\title{Effects of Physical Cooling on the Structure of Circumbinary Disks}

\author[0009-0002-8545-4492]{Minjae V. Kim}
\affiliation{Physics and Astronomy Department, Johns Hopkins University, Baltimore, MD 21218, USA}
\affiliation{{Department of Physics and Astronomy, University of Pennsylvania, 209 South 33rd St., Philadelphia, Pennsylvania, 19104, U.S.A.}}
\email[]{mjvkim@sas.upenn.edu}

\author{Julian H. Krolik}
\affiliation{Physics and Astronomy Department, Johns Hopkins University, Baltimore, MD 21218, USA}
\email{jhk@jhu.edu}

\author[0000-0003-3547-8306]{Scott C. Noble}
\affiliation{Gravitational Astrophysics Lab, NASA Goddard Space Flight Center, Greenbelt, Maryland 20771, USA}
\email[]{scott.c.noble@nasa.gov}

\begin{abstract}
{
Radiation pressure and radiative cooling often play an important role in deciding the structure and dynamics of astrophysical objects. In this paper, we present a radiative MHD simulation of a circumbinary disk (CBD) around an equal-mass SMBBH. It employs an approximate thermodynamic model based on local thermodynamic equilibrium and diffusion cooling which captures how the internal energy evolves in an optically thick region. The disk reaches a quasi-steady state within $r=5a$ whose mass distribution differs substantially from the one obtained from simulations without physical thermodynamics: the local maximum in the azimuthally-averaged surface density  and the lump are completely erased; the time-averaged inner edge becomes more eccentric; and the vertical density distribution depends strongly on the radius. These changes are attributable to the enhancement of the magnetic fields caused by the {cooling-driven} vertical compression of the disk. Our result emphasizes that the basic properties of the inner CBD are sensitive to both radiation and magnetic physics.}
\end{abstract}

\section{Introduction}
\label{sec:introduction}

Binary systems can be found in many astrophysical circumstances, incorporating stars at many different stages in their evolution as well as compact stars such as white dwarfs or neutron stars and black holes.   Because surrounding gas may accrete onto them, circumbinary disks (CBDs) are of interest in multiple contexts, from proto-stars \citep{Morii2025} to planet formation \citep{Teasdale2026} to post-asymptotic giant branch stars \citep{Huang2026} to post-common envelope stars \citep{Karino2026} to merging supermassive black hole binaries \citep{Ennoggi2026} (to avoid citing hundreds of papers, the reference illustrating each example is a very recent one). So long as the gravity binding them is Newtonian, the dynamics for all examples are very similar, with the principal contrast stemming from differing opacities within the disk gas.

This fundamental dynamical unity has led to a quasi-universal picture of their structure.  Independent of whether they employ 2D hydrodynamics or 3D magnetohydrodynamics (MHD), simulations have consistently found that when the binary mass-ratio $q \equiv M_2/M_1 \gtrsim 0.05$, a cavity is cleared out within $\sim 2a$ from the center-of-mass \citep[e.g.,][]{Artymowicz1991, Macfadyen2008, Farris2014, DOrazio2016, Shi2012, Noble2021}.
The fundamental mechanism is the destruction of closed orbits by the binary's quadrupole moment, both time-averaged and time-dependent.  Here and throughout this paper, $a$ is the binary's semimajor axis, its constant separation if the orbit is circular.   There is similar unanimity in finding that the inner edge of the CBD is very well-defined, with the surface density rising sharply toward larger radii, reaching a local maximum at $\simeq 2.5a$ {and then decreasing %
{from there to $\simeq 5a$}
{by anywhere from} a few tens of percent to a factor of few.}
\citep[e.g.,][]{Artymowicz1991, Macfadyen2008, Shi2012, Noble2012, Miranda2017, Franchini2021, Noble2021,Ennoggi+2025}. When the binary orbit is circular and $q \gtrsim 0.2$ \citep{Noble2021}, a persistent ``lump" \citep[e.g.,][]{Shi2012, Noble2012, DOrazio2013} forms at a radius near the surface density maximum.

The evolution of such systems is also unified, provided any traces of relativity are negligible.  Energy and angular momentum are exchanged between the binary and the surrounding disk.  The gravity of the circulating masses in the binary exerts a torque on the gas, simultaneously increasing both its angular momentum and its energy.  At the same time, gas that leaves the inner edge of the disk and is captured by the binary transports back to the binary both angular momentum and energy.  The characteristic scale of the gravitational torque is $\sim GM{\dot M} t_{\rm in}/a$, where $t_{\rm in}$ is the inflow time for gas in the region of strongest torque.  Because this region is a short distance closer to the center-of-mass than the nominal inner edge of the CBD, $t_{\rm in} \sim (a^3/GM)^{1/2}$.   Substituting this estimate in the previous expression yields an estimated torque $\sim (GMa)^{1/2} \dot M$.  This form is immediately recognizable as an estimate for the angular momentum brought to the binary by the accretion flow.  In other words, the two rates are automatically comparable, but have opposite sign \citep{Shi2012}.  A similar argument leads to the same conclusion for the work done by the gravitational torques and the energy given the binary by accretion.

Because the changes in the two conserved quantities governing the binary orbit, its angular momentum and energy, are both the result of close subtractions, quantitative determination of the {\it net} rate of change of the orbital elements is made very difficult. Even making a confident determination of the {\it signs} of $\dot a$ and the time-derivative of the eccentricity demands attention to all the details of the dynamics and how they may depend on system parameters.

As in many theoretical programs, initial work on CBDs adopted a number of simplifying assumptions: using 2D geometry vs. 3D; considering only hydrodynamical forces and the phenomenological $\alpha$ model of internal stress vs. magnetohydrodynamics (MHD) and physical Maxwell stress; and employing simple equations of state rather than physical heating and cooling mechanisms.  All three assumptions appeared in the very first simulational study \citep{Macfadyen2008} and continue to be used in the overwhelming majority of papers (see, e.g., \citet{Clyburn2026}). It is the goal of this paper to introduce a way to improve the situation regarding the last issue, computing disk thermodynamics in a more realistic, but computationally economical, fashion.

We are not the very first to work toward this goal. While retaining the unphysical heating of ``$\alpha$-model viscosity'', \citet{Farris2015Inspiral} suggested that cooling could be better described by thermal photon diffusion, and this approach is also used in some more recent work \citep{DeLaurentiis2025}.  However, its validity is undercut by the assumption that the internal energy is dominated by gas rather than radiation, a condition unlikely to occur in their context of interest, supermassive black hole binaries accreting at a rate capable of driving reasonably rapid orbital evolution and generating a potentially observable photon luminosity \citep{Shakura1973}.  A much more realistic scheme has been explored by \citet{Tiwari2025EqualMass, Tiwari2025UnequalMass}, in which heat released by the dissipation of MHD turbulence is radiated away, and the photon flux is calculated from a multi-angle group time-dependent transfer equation.  As we will show later, this method is, at least for the time being, limited by its computational cost, which constrains the simulation duration to only $\sim 50 - 60$ binary orbits.

In this paper, we will show that a zeroth-order approximation to radiative diffusion cooling in LTE qualitatively replicates the internal disk structure found by much more detailed calculations (as in \citet{Tiwari2025EqualMass}) but is computationally so much more efficient that durations of several hundreds of orbits can be run for a fraction of the cost of the more precise approach.   In so doing, our simulation reveals a very significant secular evolution of the disk surface density profile on the timescale of $\sim O(10^2)$~binary orbits.

\section{Method}
\label{section: method}

\subsection{Notation and units}

We use relativistic units $G=c=1$ unless explicitly stated otherwise.  Times and distances are therefore described in units of $M$, the total mass of the binary. In tensor notation, Greek indices (e.g. $\mu, \nu, \lambda$) represent the four spacetime coordinates (0,1,2,3), while Roman indices (e.g. $i,j,k$) represent the spatial coordinates (1,2,3).  We adopt the metric signature $(-+++)$.

Throughout this paper, we use shorthand notations to denote various averaging schemes. A shorthand notation for the shell average of a variable $X(r,\theta, \phi)$ is defined as 
\begin{equation}
    \shellavg{ X }
    \equiv\frac{\shellint{X}}{\int ~\dd \theta \,\dd \phi\,\sqrt{-g}}~~.
\end{equation}
where the shorthand notation for the shell integral is defined as
\begin{equation}
    \shellint{X}\equiv\int ~\dd \theta \,\dd \phi \sqrt{-g}\,X
\end{equation}
A shorthand notation for the integration of a variable $X(r, \theta, \phi)$ along the meridian is defined as
\begin{equation}
    \polarint{X}\equiv \frac{\int \dd \theta \sqrt{-g} ~ X}{\sqrt{g_{\phi \phi}}|_{\theta=\pi/2}}~~.
\end{equation}
A shorthand notation for the average of a variable $X(r, \theta, \phi)$ along the lines of latitude is defined as
\begin{equation}
    \azimavg{X} \equiv \frac{\int \dd \phi \sqrt{-g} ~X}{\int \dd \phi \sqrt{-g}}~~.
\end{equation}
A shorthand notation for time average of a variable $X(t)$ over an interval $(t_1, t_2)$ is defined as
\begin{equation}
    \timeavg{X}\equiv\frac{1}{t_2-t_2}\int_{t_1<t<t_2} \dd t ~ X(t)~~.
\end{equation}
\subsection{GRMHD Evolution}
\label{sec: GRMHD Evolution}
 The dynamics of accreting materials are described by the GRMHD equations: 
 the continuity equation
\begin{equation}
\label{eq: continuity}
    \nabla_\mu (\rho u^\mu)=0;
\end{equation}
the stress-energy conservation equation
\begin{equation}
\label{eq: stress-energy}
    \nabla_\mu T^{\mu \nu} = -\mathcal{Q} u^{\nu};
\end{equation}
and the Maxwell equations
\begin{equation}
\label{eq: maxwell}
\begin{split}
    \nabla_\mu F^{\mu \nu}&=J^\mu\\
    {\nabla_\mu } \prescript{*}{}F^{\mu \nu}&=0 .
\end{split}
\end{equation}
Here $\rho$ is the fluid rest-mass density, $u^{\mu}$ is the fluid 4-velocity, $T^{\mu \nu}$ is the MHD stress-energy tensor, $\mathcal{Q}$ is the fluid-frame cooling rate, $F^{\mu \nu}$ is the Faraday tensor, $\prescript{*}{}F^{\mu \nu}$ is the Hodge dual of the Faraday tensor, and $J^\mu$ is the electric 4-current density. Additionally, we adopt the ideal MHD condition, in which the electric field vanishes in the fluid frame:
\begin{equation}
    u_\mu F^{\mu \nu}=0~.
\end{equation}

The time evolution of these equations is calculated using the flux form of the conservation laws:
\begin{equation}
\label{eq: flux GRMHD}
    \partial_t\mathbf{U}(\mathbf{P}, \Gamma)=-\partial_i\mathbf{F}^i(\mathbf{P}, \Gamma)+\mathbf{S}(\mathbf{P}, \Gamma)
\end{equation}
where $\mathbf{P}$ is a vector of primitive variables, $\mathbf{U}$ is a vector of ``conserved'' variables, $\mathbf{F}^i$ is a vector of the fluxes, and $\mathbf{S}$ is a vector of source terms. Explicitly, these vectors are
\begin{equation}
    \mathbf{P}=\big(\rho, u, \tilde{u}^i, B^i\big)^{T}
\end{equation}
\begin{equation}
    \mathbf{U}(\mathbf{P}, \Gamma)=\sqrt{-g}~\big(\rho u^t, {T^t}_t+\rho u^{t}, {T^t}_j, B^k\big)^{T}
\end{equation}
\begin{equation}
    \mathbf{F}^i(\mathbf{P}, \Gamma)=\sqrt{-g}~\big(\rho u^i, {T^i}_t+\rho u^{i}, {T^i}_j, b^iu^k-b^ku^i\big)^{T}
\end{equation}
\begin{equation}
\label{eq: source term}
    \mathbf{S}(\mathbf{P}, \Gamma)=\sqrt{-g}~\big(0, {T^\mu}_\nu{\Gamma^{\nu}}_{t\mu}-\mathcal{Q}u_t, {T^\mu}_\nu{\Gamma^{\nu}}_{j\mu}-\mathcal{Q}u_j, 0\big)^{T},
\end{equation}
where $g$ is the determinant of the metric $g_{\mu, \nu}$, $B^i= \prescript{*}{}{F^{it}}/\sqrt{4\pi}$ is the magnetic field, and  $b^{\mu}=u_\nu \prescript{*}{} F^{\nu\mu}$ is the magnetic 4-vector. The primitive velocity $\widetilde{u}^i$ is defined as the flow's velocity projected onto a frame moving orthogonal to the space-like hypersurface:
\begin{equation}
    \widetilde{u}^i=u^i-u^tg^{ti}/g^{tt}.
\end{equation}

The MHD stress-energy tensor is the sum of the fluid stress-energy tensor and the electromagnetic stress-energy tensor. The fluid stress-energy tensor is 
\begin{equation}
\label{eq: stress-energy matter}
    T^{\mu\nu}_{\mathrm{fluid}}=\rho hu^{\mu}u^{\nu}+pg^{\mu \nu},
\end{equation}
where $h=1+\epsilon+p/\rho$ is the specific enthalpy, $p$ is the pressure of the matter, and $\epsilon$ is the specific internal energy density. The electromagnetic stress-energy tensor is
\begin{equation}
    T^{\mu \nu}_{\mathrm{EM}}=2 p_{\mathrm{mag}} u^{\mu}u^{\nu}+ p_{\mathrm{mag}}g^{\mu\nu}-b^\mu b^\nu
\end{equation}
where $p_{\mathrm{mag}}=b^\mu b_\mu/2$ is the magnetic pressure. We close the fluid equations by assuming the EOS
\begin{equation}
    p = (\Gamma-1)u
\end{equation}
where $u=\rho \epsilon$ is the internal energy density in the fluid frame and $\Gamma$ is the local effective adiabatic index. During the GRMHD evolution step, $\Gamma$ is assumed to be constant.
 
 To solve the GRMHD equations, we use the intrinsically-conservative GRMHD code HARM3D \citep{Gammie2003, Noble2006, Noble2009}. Unlike
 previous circumbinary disk simulations with HARM3D, the effective adiabatic index can vary. However, the local conservation equations are agnostic to derivatives of the adiabatic index, so no change to their numerical implementation in HARM3D is required. Further details about HARM3D can be found in \cite{Noble2009, Noble2012}.

Relativistic dynamics require specification of a metric.  To represent the metric created by an equal-mass black hole binary in a circular orbit, we use a 2.5-order post-Newtonian (PN) expansion \cite{Mundim2014} in its near-zone limit \cite{Noble2012, Noble2021}.  Validity and rapid convergence of this expansion are guaranteed by the scales we study, which are never closer to the binary's center-of-mass than $75M$ and extend outward to $5000M$.  We choose this range of radii because, in units of the binary semimajor axis $a$, it corresponds to $0.75 - 50 a$.  The associated binary orbital period is $t_{\mathrm{bin}}=6.37 \times 10^3M$. 

\subsection{Thermodynamic Model}
\label{sec: Thermodynamic Model}

{As we remarked earlier, nearly all CBD simulations to date have used very simplified thermodynamics for both the cooling rate and the EOS. Cooling in most studies has been represented either by an {\it ad hoc} target-temperature \citep{Wang2023} or target-entropy \citep{Noble2012} form with an arbitrarily specified cooling rate, or entirely omitted \citep{Macfadyen2008, Shi2012}. A few hydrodynamics studies (e.g.,\citet{Farris2015Thermal, Cocchiararo2024}) adopted a nominally physical cooling rate ($\propto T^4$) meant to capture diffusive thermal emission from the disk surface, but these, too, rest on two major simplifications: that the heat is generated by viscous dissipation { at a rate proportional to the local pressure and} that the local diffusion time depended only on the opacity and the local surface density, making no allowance for irregularities in the density distribution.
Moreover, both $\alpha$-hydro and MHD simulations --- with the exception of \citet{Tiwari2025EqualMass, Tiwari2025UnequalMass, Cocchiararo2026} --- have ignored radiation pressure and assumed either an isothermal EOS or a  $\Gamma$-law EOS  (i.e., $p = (\Gamma - 1)u$) for some constant $\Gamma$ (usually either $\Gamma = 4/3$ or $\Gamma = 5/3$).}

Physically-based cooling is difficult to treat well because its character depends strongly on the optical depth $\tau$, both directly through the escape rate and indirectly through the thermal decoupling of gas and photons.  Our approach interpolates both effects between the extremes of $\tau \gg 1$ and $\tau \ll 1$. In the limit of large optical depth, the fluid is assumed to be in perfect local thermodynamic equilibrium with the photons, and the escape time is given by the customary diffusion expression: $t_{\rm diff} = \tau h/c$, where $h$ is a characteristic lengthscale for the density distribution. On the other hand, in the limit of small optical optical depth, the photon energy density decouples from the gas energy density, becoming a small fraction of the LTE limit, and the time required for photons to leave a region is always $\simeq h/c$. 

To be specific, for a given gas temperature $T$, we approximate the photon energy density by $f_{\rm eff}(\tau)a_{\rm SB}T^4$, with
\begin{equation}
\label{eq: feffective}
    f_{\mathrm{eff}}(\tau)=e^{-c_1/\tau}\frac{\tau^{c_3}}{c_2+\tau^{c_3}}
\end{equation}
and $c_1=10$, $c_2=1000$, and $c_3=10$. These parameters were chosen to ensure that $f_{\rm eff}$ is very close to unity wherever the optical depth is high, but small enough to make gas pressure dominate radiation pressure where the optical depth is $ \lesssim 1$.   
The other interpolation is in the diffusion time, which we write as
\begin{equation}
\label{eq: diffusion time}
t_{\rm diff} = \frac{(\tau + 1)h}{c},
\end{equation}
but with special definitions of both $\tau$ and $h$.

The gas temperature can then be determined by solving
\begin{equation}\label{eq: internal energy}
u = (3/2) \rho k_BT/\bar{m} + f_{\rm eff}(\tau) a_{\rm SB} T^4,
\end{equation}
where $\bar{m}$ is the mean mass per particle (see Appendix~\ref{sec: EOS temperature appendix} for details).
The cooling rate is then given by $\mathcal{Q} = f_{\rm eff}(\tau) a_{SB}T^4/t_{\rm diff}(\tau)$.  With the temperature, we can also immediately evaluate the local effective adiabatic index
\begin{equation}
\label{eq: eff adiabatic index}
    \Gamma = \frac{u_{\mathrm{gas}} \Gamma_{\mathrm{gas}}+u_{\mathrm{ph}}\Gamma_{\mathrm{ph}}}{u_{\mathrm{gas}}+u_{\mathrm{ph}}}\\
\end{equation}
where $u_{\mathrm{gas}}$ is the internal energy density of the gas, $\Gamma_{\mathrm{gas}}=5/3$ is the adiabatic index of the gas, $u_{\mathrm{ph}}$ is the internal energy density of the photons, and $\Gamma_{\mathrm{ph}}=4/3$ is the adiabatic index of the photons.

To find the local optical depth, we follow a procedure outlined in \citet{MurguiaBerthier2021}.  
The goal of this procedure is to find the path from each cell in the simulation to the outside boundary along which the optical depth is minimal.  Our method begins by assigning each cell a tentative minimum optical depth $\tau$.  We then begin a series of iterations.  In each iteration, we pass through the array of cells, updating their values of $\tau$; this pass touches each cell exactly once.
The update follows the rule 
\begin{equation}
    \tau(n+1)=\min \big(\tau_{\mathrm{neighbor}}(n)+\kappa_T \overline{\rho} (\overline{g_{ij}}\dd x^i \dd x^j)^{1/2}\big),
\end{equation}
where $\tau(n+1)$ is the tentative minimum optical depth for the cell in question in the $n+1$th iteration and its six nearest cells qualify as ``neighbors''.  $\bar{\rho}$ is the mean of this cell's and its neighbor cell's density; $\overline{g_{ij}}$ is the mean metric element in the same sense. The key assumption of this prescription is that if the path from your cell to the outside passes through a particular neighbor cell, the path's total optical depth to the outside must be the sum of the neighbor's optical depth and the amount incurred moving from the center of the cell of interest to the center of the neighboring cell.  These iterations are repeated until none of the $\tau$ values in the $n+1$th iteration is significantly different from its value in the $n$th iteration.
The optical depths found after the iterations converge are then used in Eq.~\ref{eq: diffusion time}, along with an estimated value for the length of the minimal path from each cell to the outside,
\begin{equation}
h = \tau/(\kappa_T \rho),
\end{equation}
where $\rho$ is the density at the location whose optical depth is $\tau$.
{Because the region outside the photosphere adds very little to the minimal path integral when it starts from an optically thick point, it makes very little difference whether the endpoint is the photosphere or the edge of the problem volume; we choose the latter.}

Because the hydrodynamic timestep is typically much smaller ($\sim 0.1 M$) than the cooling timescale ($\sim 10^4 M$), we update $\tau$ and $\mathcal{Q}$ once every 10 timesteps. However, in regions near the photosphere, the cooling timescale may be significantly shorter ($\sim 1M$). To minimize the error from using longer intervals between the cooling rate updates, we use a time-averaged cooling rate $\timeavg{\mathcal{Q}}$ over the time difference $\Delta t$ between reevaluations of $\tau$ and $\mathcal{Q}$.  An analytic solution for the mean cooling rate is
\begin{equation}
    \timeavg{\mathcal{Q}}=\frac{u}{\Delta t}\Bigg(\frac{Y^{-1/3}(Y^{-1}+4c_1)}{4c_1 + 1}-1\Bigg)
\end{equation}
where 
\begin{equation}
    c_1 = \frac{1}{4}\frac{\Gamma_{\mathrm{gas}}-\Gamma}{\Gamma-\Gamma_{\mathrm{ph}}}, \quad c_2 = \frac{3}{4}\frac{\kappa \rho c \Delta t}{\tau^2}
\end{equation}
\begin{equation}
    Y = W(c_1 e^{c_1 + c_2})/c_1.
\end{equation}
Here, the Lambert-W function $W(x)$ is defined as the solution of
\begin{equation}
    W(x)e^{W(x)}-x=0.
\end{equation}
A detailed derivation of these results is provided in Appendix \ref{sec: Lambert-W function appendix}.

\subsection{Simulation Details}
\label{sec: Simulation Details}

To explore the sensitivity of circumbinary disk structure to cooling physics, we contrast a simulation run with our new disk thermodynamics procedure (DiffRUN) with another (LegacyRUN) that differs in two ways: it assumes $\Gamma = 5/3$ everywhere and it employs a target-entropy cooling rate, i.e., one in which
\begin{equation}
\label{eq: target entropy function}
    \mathcal{Q}=
    \begin{cases}
        \frac{u \sqrt{2}}{t_{\mathrm{orb}}}\left(\frac{S}{S_0}-1\right)^{1/2}& \text{$S>S_0$}\\
        0 & \text{$S\leq S_0$}.
    \end{cases}
\end{equation}
Here, $t_{\mathrm{orb}}$ is the period of a circular Keplerian orbit at the local radius, $S=(p/\rho^{5/3})/(c^2\rho_0^{-2/3})$ is the local entropy proxy, and $S_0$ is the target entropy proxy.  Although they differ in their cooling rates and adiabatic indices, they are run with versions of HARM3D that are identical otherwise. To ensure that they are solving the same problem, the initial data for DiffRUN are taken from a stage of LegacyRUN in which it has achieved statistical steady-state with inflow equilibrium out to {$r \simeq 5a$}.

LegacyRUN was initialized using the hydrostationary torus solution, like in \cite{Noble2012}, which follows \citet{Chakrabarti1985} and \citet{DeVilliers2003}, to determine the density and pressure distribution.  We began by setting $a=100M$. 
Because we expect the inner edge of the CBD to be $\gtrsim 2a$, we placed
the torus's inner edge at $r_{\mathrm{in}}=3a$, its pressure maximum at $r_{\mathrm{p}}=6.8a$, and its outermost extent at $r_{\mathrm{out}}\simeq 40a$, as in Run$_\mathrm{lrg}$ of \cite{Noble2021}. The code-unit of density is unity at the density maximum. The initial disk is isentropic with $S_0=2.47\times 10^{-5}$, which we use as the target entropy for Eq. \ref{eq: target entropy function}.

The initial solution is tuned to have an aspect ratio of $H/r=0.1$ at $r=r_{\mathrm{p}}$, where the azimuthally averaged density scale height $H(r)$ is 
\begin{equation}
    H(t,r)=\frac{\shellavg{\rho \sqrt{g_{\theta \theta}}\left|\theta-\pi/2\right|}}{\shellavg{\rho}}.
\end{equation}

A poloidal magnetic field contained inside the disk is derived from a vector potential
\begin{equation}
    A^{\phi}=A_0 \max \bigg[\bigg(\rho - \frac{1}{4}\rho_{\mathrm{max}}\bigg),0\bigg]
\end{equation}
where $A_0$ is set so that the disk's total internal energy is 100 times its total magnetic energy.
Two numerical floors are imposed: $\rho_{\mathrm{floor}}=1.62 \times 10^{-6} (r/M)^{-3/2}$ and $u_{\mathrm{floor}}=1.62 \times 10^{-8} (r/M)^{-5/2}$.

We deemed LegacyRUN to have reached a rough equilibrium after $\sim 400 t_{\rm bin}$.  Our initial data for DiffRUN were then constructed from a snapshot of LegacyRUN at $t\simeq 417 t_{\mathrm{bin}}$.

Although a simulation with only gas pressure can be run with an arbitrary unit of density, this is no longer true when radiation is involved because the opacity introduces a specific scale for density.  In order for DiffRUN to maintain an aspect ratio $H/r \simeq 0.1$ and a ratio of cooling time to orbital time $\simeq 2$ at $r=3a = 300M$, we redefined the density code-unit to be $\rho_0 = 1.0 \times 10^{-11}$~gm~cm$^{-3}$. This choice implies an accretion rate in Eddington units $\dot{m} \sim 20$, assuming a radiative efficiency $\eta = 0.1$. 

For these values, the solution of Eq. \ref{eq: internal energy} leads to a situation in which the radiation pressure is much larger than the gas pressure. This fact may have hydrodynamic consequences when transitioning from LegacyRUN to DiffRUN
{because, for fixed internal energy $u$, a radiation-dominated fluid has only half the pressure of a gas-dominated fluid.  To avoid numerical instabilities as a consequence of this change, we break it into two parts.  First we change $\Gamma$ from 5/3 to 4/3, doubling the internal energy so as to preserve the pressure, but continuing to interpret the pressure as entirely due to gas.  This change also has the consequence of increasing $S$ for nearly all the mass of the disk, so cooling is sharply increased. We then continue the run for $2t_{\rm bin}$.  At this point, we complete the transformation by introducing radiation energy into the definition of internal energy and substituting our new diffusion-based cooling procedure for the old target-entropy version.} We also make slight adjustments to the numerical floors: $\rho_{\mathrm{floor}}=2.00 \times 10^{-6}(r/M)^{-3/2}$ and $u_{\mathrm{floor}}=4.00\times 10^{-8}(r/M)^{-5/2}$. 

The snapshot at the end of the buffer run is the initial condition of DiffRUN.  Its time is $t_0 \simeq 419 t_{\rm bin}$; we describe times within DiffRUN in terms of $t-t_0$. The buffer run is excluded from any results we present. 

To provide a longer run of comparison data, we continued LegacyRUN until $t\simeq 540t_{\mathrm{bin}}$ or $t-t_0 \simeq 121t_{\mathrm{bin}}$ while simulating DiffRUN until $t-t_0 \simeq 382 t_{\mathrm{bin}}$.

Both LegacyRUN, buffer run and DiffRUN use the same grid scheme as Run{$_{\rm lrg}$} of \cite{Noble2021}, with $420 \times 160 \times 400$ cells in spherical coordinates $(r, \theta, \phi)$. The radial coordinates of the cells increase exponentially from $r=0.75a=75M$ to $r=50a=5000M$, maintaining a fixed ratio $\Delta r / r$.  The azimuthal coordinates are equally spaced. The polar angle ($\theta$) coordinates are defined through a polynomial function defined in Eqn.~22 of \cite{Noble2012} that creates higher resolution near the midplane than near the polar axis. The region within $0.2$~radian from the axis is excluded from the grid. We use $0^\mathrm{th}$-order extrapolation or outflow boundary conditions at the radial boundaries. Reflective  conditions at the polar boundaries are used to minimize outflows/inflows; there we set the velocity's and magnetic field's $\theta$-component to be anti-symmetric and mirror copy the nearest physical cell values into the ghost zones. {The disks satisfied the MRI quality conditions of \cite{Hawley2011, Hawley2013}, which we demonstrate in App. \ref{sec: MRI resolution}}.

\section{Results}
\label{sec:results}

\subsection{Mass Inflow}
\label{sec: mass inflow}

The mass inflow timescale in our simulation ($\mathcal{O}(100) t_{\text{bin}}$ at $r \simeq 4a$) is orders of magnitude greater than the thermal timescale ($\gtrsim t_{\text{bin}}$ at $r \simeq 4a$) or the %
dynamical timescale ($\sim t_{\text{bin}}$) at $r \simeq 4a$). Therefore, at any given time, the thermal and dynamical states are in an equilibrium consistent with the roughly time-steady surface density profile. By running for several inflow timescales, we can both establish an approximate inflow equilibrium at $r \lesssim 4a$, and detect any slow secular evolution of that equilibrium.

\begin{figure}[thbp!]
    \centering
    \includegraphics[width=\columnwidth]{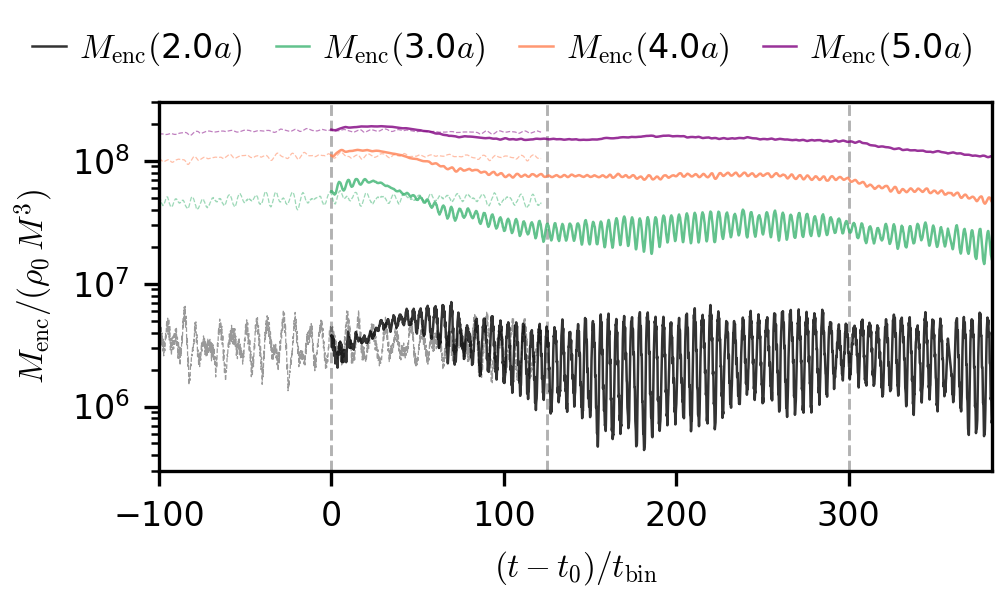}
    \caption{Mass enclosed $M_{\rm enc}$ within various radii as functions of time. Presented on a logarithmic scale. Each vertical line is a boundary between epochs. (Solid) DiffRUN, (dashed) LegacyRUN.}
    \label{fig: mass enclosed}
\end{figure}

{Whether the mass enclosed within a given radius is constant over time is a direct test of inflow equilibrium.  This quantity is shown in Figure \ref{fig: mass enclosed} for several radii in both DiffRUN and LegacyRUN.
We define the enclosed mass by}
\begin{equation}
	M_{\text{enc}}(t,r) = \int_{r^\prime < r} \dd r^\prime ~ \dd \theta~ \dd \phi~ \sqrt{-g}~ \rho~.
\end{equation}

Two sorts of time-dependence can be seen. There is secular evolution leading to a roughly time-steady state, and there is also a high-frequency modulation.  The latter is frequently seen in CBD simulations and stems from a periodicity in the mass accretion rate across the disk's cavity driven by the binary's time-dependent gravitational torque. Its only interest in the present context is that a comparison between DiffRUN and LegacyRUN shows that it has a larger amplitude in DiffRUN.  The former is the principal result of this paper.
Although $M_{\text{enc}}$ in LegacyRUN remains flat from $t-t_0=-100 t_{\text{bin}}$ until the end of its run, DiffRUN experiences secular changes in this quantity {signaling a redistribution of matter within the disk as a result of}
the introduction of our physical thermodynamic model.

It is convenient to use the secular evolution of ${M}_{\text{enc}}(t, r)$ as a criterion for division of DiffRUN into three epochs. At all four radii shown in Figure~\ref{fig: mass enclosed}, ${M}_{\text{enc}}(t, r)$ increases and then decreases during Epoch 1 ($t-t_0 \leq 125 t_{\text{bin}}$), and plateaus during Epoch 2 ($125 t_{\text{bin}} \leq t-t_0 \leq 300 t_{\rm bin}$).  In Epoch 3 ($t-t_0 \geq 300 t_{\text{bin}}$), the mass enclosed decreases slowly in the three outer regions, but remains nearly constant inside $2a$. Each epoch has distinct physical characteristics, which we discuss in the following sections.
\begin{figure}[thbp!]
    \centering
    \includegraphics[width=\columnwidth]{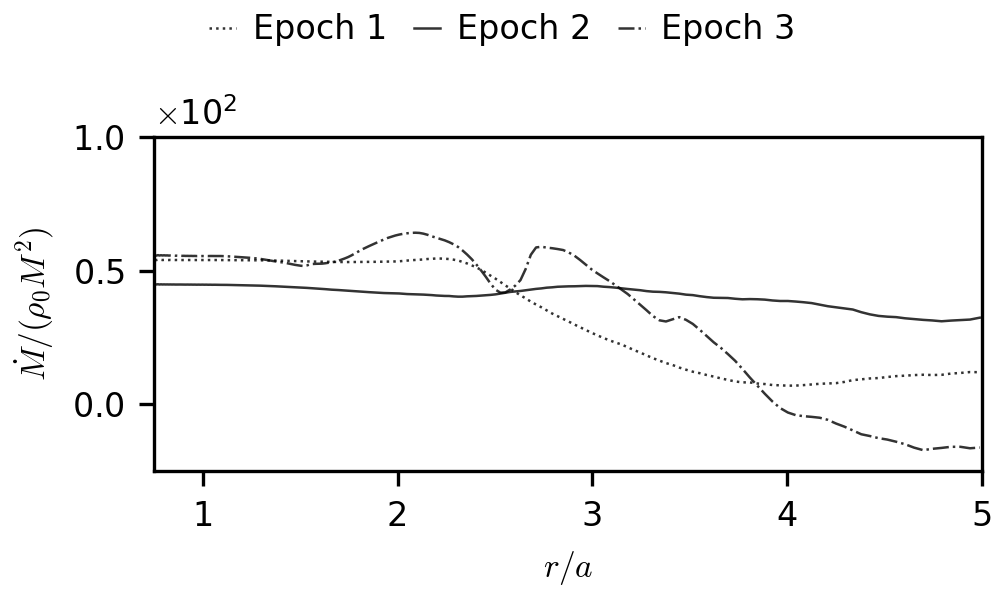}
    \caption{DiffRUN's mass inflow rate $\dot {M}$ as a function of radius, time averaged over each epoch.}
    \label{fig: mdot epochs}
\end{figure}

Constancy with radius of the time-averaged mass inflow rate $\timeavg{\dot M}$ tests the quality of inflow equilibrium in a different way. The instantaneous mass inflow rate is defined as
\begin{equation}
     \dot M(t,r) \equiv\int \, \dd \phi \, \dd \theta \sqrt{-g} \rho u^r~~.
\end{equation}

In a perfect steady state, the gradient $\partial_r\dot M$ vanishes. The path that DiffRUN takes to evolve toward mass inflow equilibrium can be seen in Figure~\ref{fig: mdot epochs}, which displays $\dot M$ averaged over each of three epochs. During the first two epochs, the range of radii over which $\dot M(r)$ is nearly constant grows, extending from $\approx 2.5a$ in Epoch~1 to $\gtrsim 5$ in Epoch~2. However, in Epoch~3, the edge of inflow equilibrium retreats to $r \simeq 3.5a$. In Epoch~1, the sense of mass motion implied by the radial gradient of $\dot M$ is that mass drains from $2.5a \lesssim r \lesssim 4a$ and passes through the region $2.5a \lesssim r < 3.5a$ before being accreted onto the binary. In Epoch~2, there is little net mass-transfer anywhere inside $\simeq 5a$.  Finally, in Epoch~3, there is net mass loss in the region $3a \lesssim r \lesssim 4a$, but, as in Epoch~1, this mass passes through the smaller radii of the CBD and cavity in order to join the binary. 

The mass inflow rate measured at the inner radial boundary of the simulation region ($\dot M (t, r_{\rm in})$) gives a good estimate of the binary's accretion rate. During the quasi-steady state of LegacyRUN it is %
$\dot{M} \simeq 44 \rho_0 M^2$ (averaged over the last $120 t_{\rm bin}$). With the introduction of the photon diffusion model, the accretion rate varies up and down by several tens of percent, but is generally $\sim 20\%$ greater (see Fig.~\ref{fig: mdot epochs}).
During Epoch 3, the binary, on average, accretes at a rate similar to the quasi-steady state of LegacyRUN. 

\subsection{Shell-Integrated Structure}
In this section, we describe the evolution of shell-integrated disk structures.  To do so we focus on the azimuthally-averaged surface density, calculated as
\begin{equation}
    \Sigma(t,r) = 
  \frac{ \shellint{ \rho } }{\int d\phi \left.\sqrt{g_{\phi \phi}}\right|_{\theta=\pi/2}}~.
\end{equation}

\subsubsection{Elimination of Surface Density Maximum}
\label{sec: Azimuthally-Averaged Surface Density}

\begin{figure*}[thbp!]{
    \centering
    {
    \includegraphics[width=2\columnwidth]{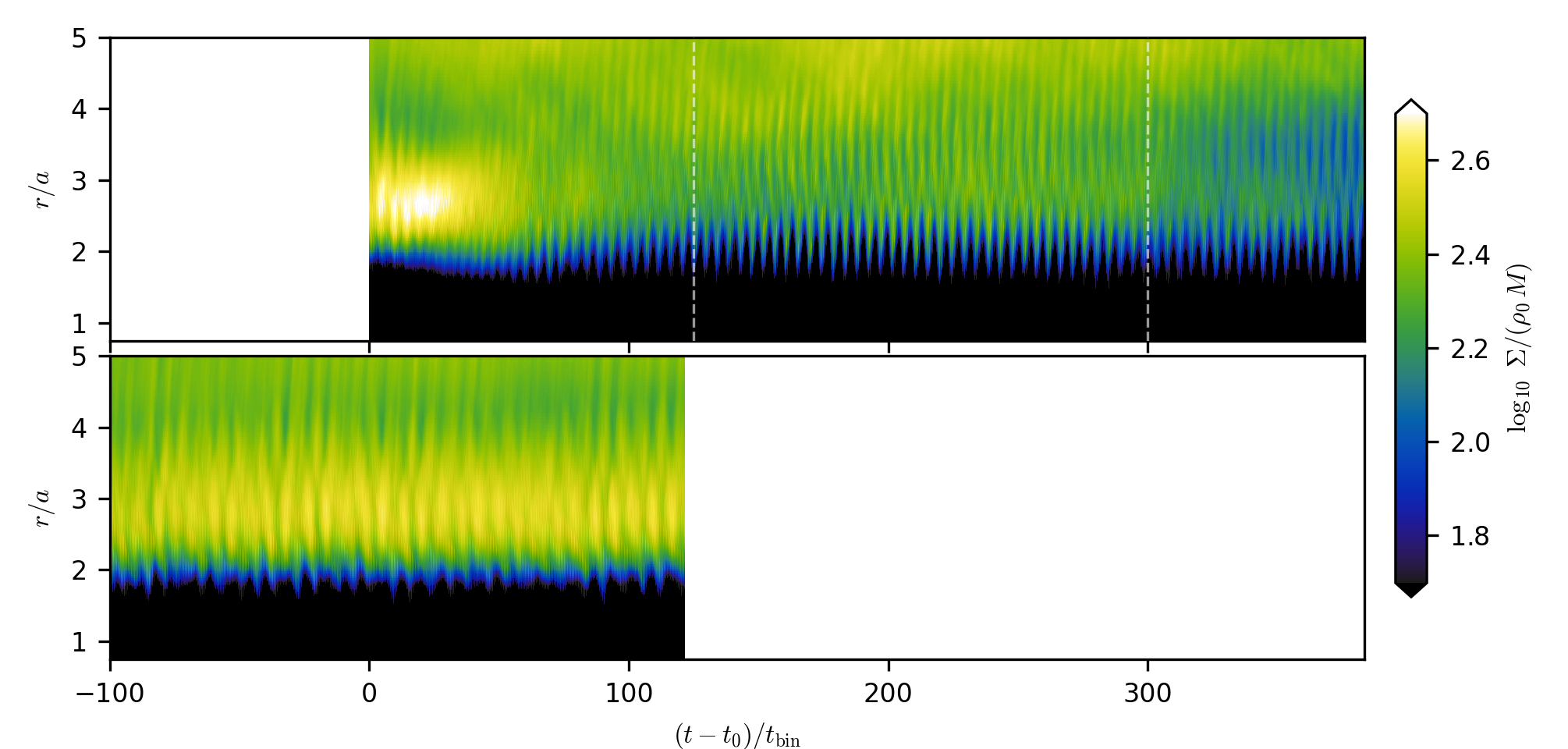}
    }
    \caption{Color contour of azimuthally-averaged surface density as a function of radius and time. Presented in a logarithmic scale. Each vertical line is a boundary between epochs. (Top) DiffRUN, (bottom) LegacyRUN.}
    \label{fig: surface density time}
}
\end{figure*}

\begin{figure*}[thbp!]{
    \centering
    {
    \includegraphics[width=2\columnwidth]{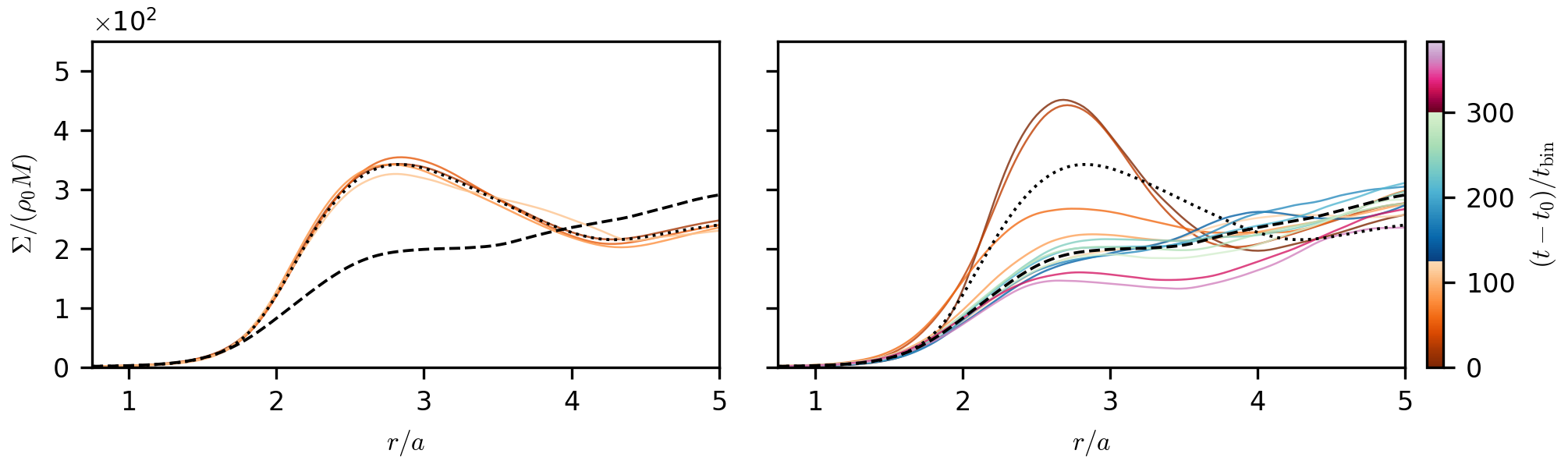}
    }
    \caption{Azimuthally-averaged surface density $\Sigma$ as a function of radius at various times. Each line is box-car smoothed over $30t_{\rm bin}$. Colorbar on the right indicates the time that each line corresponds to, which progresses as follows: dark brown to light orange (Epoch 1), dark blue to light green (Epoch 2), dark red to light purple (Epoch 3). The black dashed lines are the average of DiffRUN over Epoch 2 ($125 t_{\mathrm{bin}}<t-t_0<300t_{\rm bin}$). The black dotted lines are the average of LegacyRUN over the last $175t_{\rm bin}$($-54t_{\rm bin} < t-t_0 < 121t_{\rm bin}$). (Left) LegacyRUN, (right) DiffRUN.} 
    \label{fig: surface density time smoothed}
}
\end{figure*}

For both DiffRUN and LegacyRUN, Figure~\ref{fig: surface density time} shows the time evolution of $\Sigma(t,r)$ %
{in a spacetime format}. Figure~\ref{fig: surface density time smoothed} shows the secular evolution of $\Sigma(t,r)$ for DiffRUN and LegacyRUN utilizing a series of time-smoothed line plots.

{Like $M_{\rm enc}(t,r)$, a quantity to which the surface density is closely related, $\Sigma(t,r)$ in both DiffRUN and LegacyRUN display both short-period modulation and long-term secular evolution.  In one respect, the shape of $\Sigma(r)$ in both simulations resembles the pattern found in all previous CBD simulations (for example, \cite{Macfadyen2008, Shi2012, Noble2021} among many others) in that the disk has a sharply cut-off inner edge at $r \sim 2a$.}

{However, there is also a noteworthy change. In LegacyRUN, the shape of $\Sigma(r)$ closely mimics that found in essentially every previous paper: it has a prominent local maximum at $r \simeq 2.5a$.  In sharp contrast, by the time DiffRUN has evolved for $\sim 150 t_{\rm bin}$, the surface density maximum has completely disappeared, replaced by a gradual rise in surface density with increasing radius that runs from the cavity out to $r \gtrsim 5a$, but with a flattening between $r \simeq 2.5a$ and $3.5a$.}

Interestingly, the evolutionary path toward this result is indirect. For the first $\sim 30 t_{\rm bin}$, fluid accumulates in the region $2.0 a\lesssim r \lesssim 3.5 a$, and the local maximum becomes {\it more} prominent, peaking at a value $\simeq 20\%$ higher than in LegacyRUN. For the remainder of Epoch~1 ($30 - 125 t_{\rm bin}$), the outermost part ($r >4a$) changes very little, and the innermost part ($r < 2a$) changes slightly, its slope becoming somewhat gentler. However, the local maximum erodes so sharply that, by $t \simeq 120t_{\rm bin}$, it is no longer identifiable. Subsequently, $\Sigma(r)$ maintains a nearly stable profile during Epoch~2 ($125 - 300 t_{\rm bin}$).  The only noticeable alteration is a small increase followed by a return to its original value in the radial range $4 - 5a$. Secular evolution resumes during Epoch~3.  {Although the plateau at $2 - 3.5a$ remains, it falls in magnitude by $\simeq 30\%$. Similarly, the surface density at $r >4a$ also falls, but by a fractional amount roughly half of the drop in the plateau value.}

To sum up, although a surface density maximum near the CBD's inner edge has long been a staple feature of CBD simulations using simplified thermodynamics, changing to a more physical cooling rate and including radiation pressure in the equation of state (EOS) lead to the maximum's complete erasure.  It is possible that further secular evolution introduces new features, but the rate of change for $t \gtrsim 125 t_{\rm bin}$ is much smaller than during the first third of the simulation.

\subsubsection{Evolution of Shell-Integrated Magnetic Properties}
\label{sec:shell-integrated magnetic}

\label{sec: Evolution Magnetic}
\begin{figure*}[thbp]
    \centering
    \includegraphics[width=2\columnwidth]{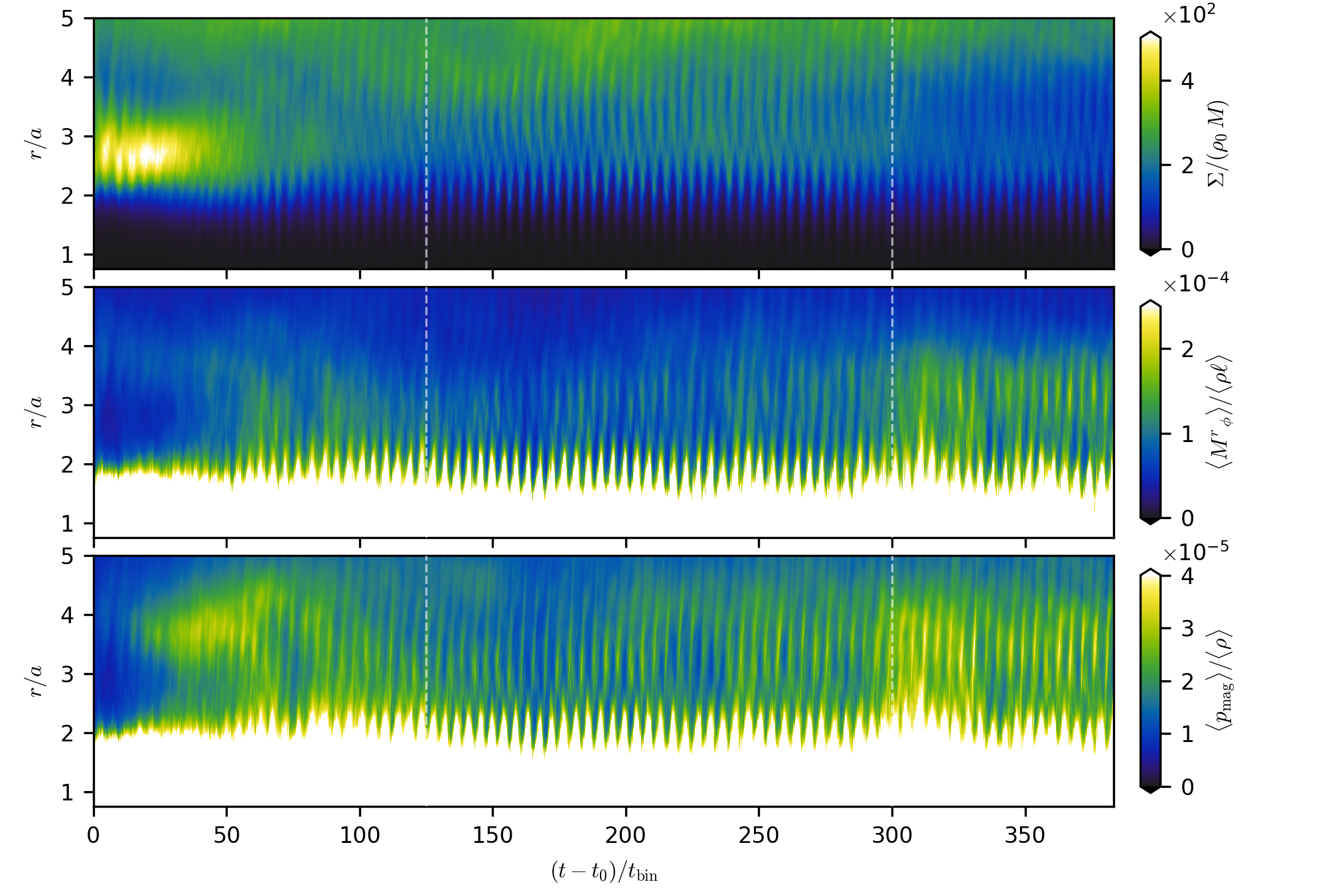}
    \caption{Color contours {on a linear scale} of various shell averaged quantities as functions of radius and time in DiffRUN. 
    (From top to bottom) azimuthally-averaged surface density $\Sigma$; the ratio of shell-averaged Maxwell stress to shell-averaged angular momentum {density} $\shellavg{M^r {}_\phi}/\shellavg{\rho \ell}$; the ratio of shell-averaged magnetic pressure to shell-averaged rest-mass density $\shellavg{p_{\rm mag}}/\shellavg{\rho}$.  
    Each vertical line is a boundary between epochs. 
    }
    \label{fig: comparison of development}
\end{figure*}

The radial redistribution of mass is dictated by how internal stresses transport angular momentum. In the bulk of the disk ($r \gtrsim 2a$), the shell-averaged Maxwell stress $M^r {}_\phi= 2p_{\mathrm{mag}}u^ru_{\phi}-b^{r}b_{\phi}$ is roughly $3$ times the Reynolds stress $R^r {}_\phi=\rho h ~\delta u^r~\delta u_\phi$.  This ratio of Maxwell to Reynolds stress is consistent with the results from previous MHD simulations of CBDs \citep{Shi2012, Noble2021}. We therefore focus on the evolution of magnetic stress and magnetic pressure. Displayed in Fig.~\ref{fig: comparison of development} are: {the surface density;} the ratio of Maxwell stress to fluid angular momentum{\footnote{In the code's output the specific angular momentum is defined as $\ell \equiv u_\phi/u_t$; the ratio displayed is the inverse of the time over which magnetic stress can carry a fluid element's angular momentum across the fluid element}} $\shellavg{M^r {}_\phi}/\shellavg{\rho \ell}$ and the ratio of magnetic pressure to density $\shellavg{p_{\rm mag}}/\shellavg{\rho}$. This plot thus emphasizes the change in magnetic field's ability to transport angular momentum and accrete mass. Alternative perspectives on change in magnetic field's dynamic significance, in terms of $p_{\rm mag}$ and plasma $\beta$, are provided in Figs. \ref{fig: average density vertical} and \ref{fig: vertical support}.

Like $\Sigma(t,r)$, $\shellavg{M^r {}_\phi}/\shellavg{\rho \ell}$ and $\shellavg{p_{\rm mag}}/\shellavg{\rho}$ display short-period modulations and long-term secular evolution. The short-period modulations, with frequency $\simeq 0.2 \Omega_{\rm bin}$, are driven by density waves {excited by the time-dependent quadrupole moment of the binary.} %
They contribute only marginally to mass redistribution, %
so we focus on secular evolution for the rest of the section; {we will, however, discuss the short-period modulations in Sec.~\ref{sec: Temporal Modulations}.}

A salient feature in the secular development of 
{both the ratio of stress to angular momentum and the ratio of magnetic pressure to density is that both have notably higher values for $2a \lesssim r \lesssim 4a$ during two extended periods of time, one near the beginning of the simulation and one near the end.
Where the contrast is greatest ($r \simeq (2.5 - 3)a$), during the first episode both measures of magnetic field strength reach values as high as $\approx 3\times$ their initial values.
From $t-t_0 \simeq  70t_{\rm bin}$ until $t-t_0 \simeq 300 t_{\rm bin}$, both quantities drop back in magnitude, but remain $\sim 2 \times$ their initial values.
At $t-t_0 \simeq 300 t_{\rm bin}$, both rise again, reaching even higher levels, $\sim 3 - 4 \times$ the initial state.}

Strikingly, both magnetic enhancements are followed by a significant decrease in $\Sigma(r)$ in
{exactly the region where the enhancements occur.
During the first episode, the surface density at $r \simeq (2 - 3)a$ begins to fall within a few orbits of when the magnetic pressure begins to rise, and the surface density in this range of radii levels out around the time when the two measures of magnetic effects begin to decline.}
{The tight connections, both temporal and spatial,} 
between the magnetic enhancements and draining of mass in the region suggest a causal relationship between the two. This point will be discussed further in Sec. \ref{sec: discussion}.

\subsubsection{Evolution of Shell-Integrated Thermal Properties}
\label{sec: evolution of shell-integrated thermal properties}

\begin{figure*}[htbp]
    \centering
    \subfloat{
    \includegraphics[width=2\columnwidth]{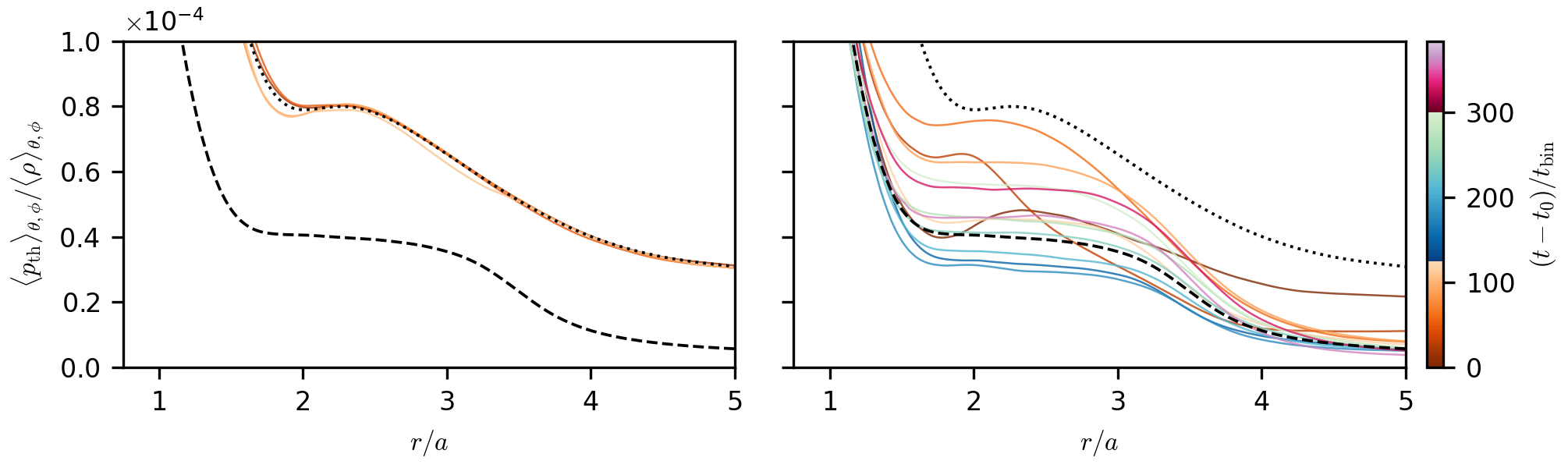}
    }
    \\[-16pt]
    \subfloat{
    \includegraphics[width=2\columnwidth]{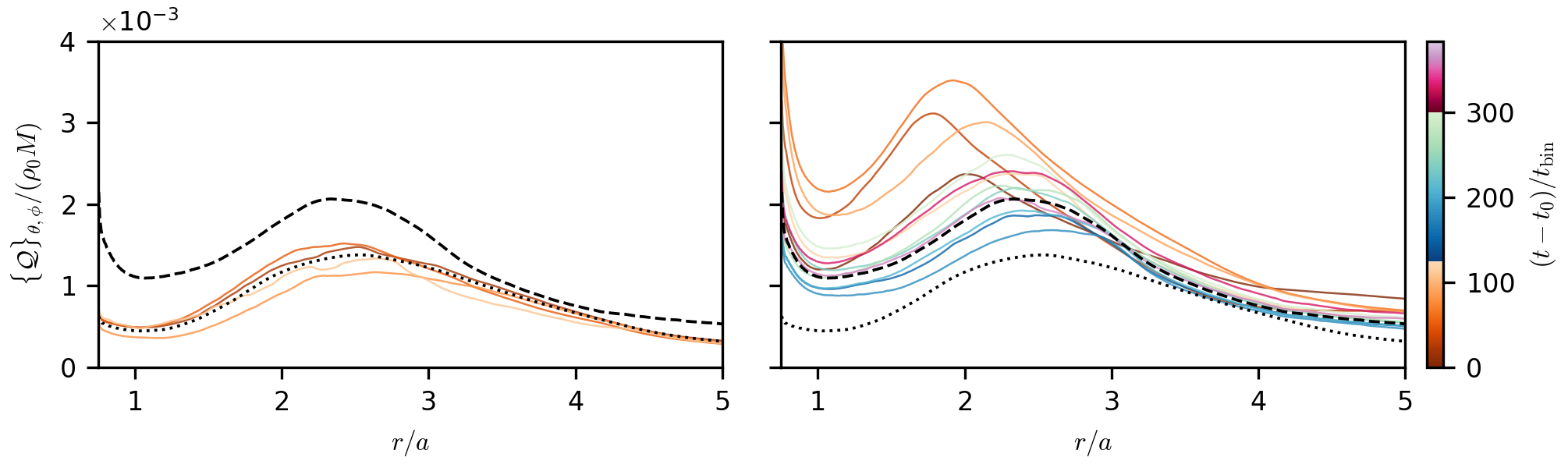}
    }
    \\[-16pt]
    \subfloat{
    \includegraphics[width=2\columnwidth]{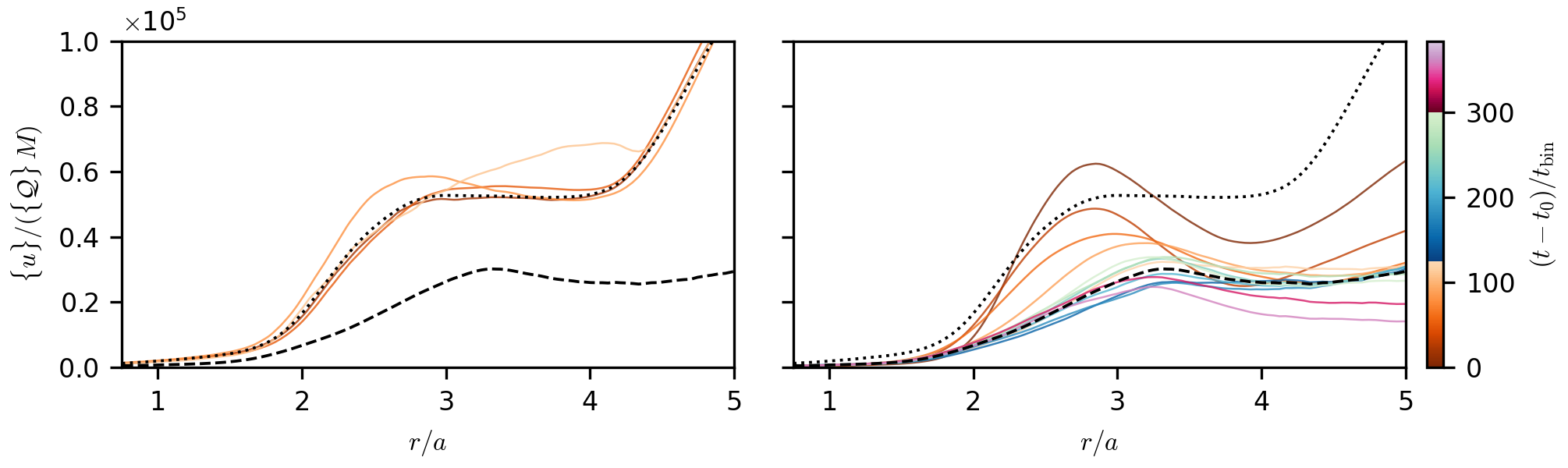}
    }
    \caption{Shell-integrated thermal properties, box-car smoothed over $30t_{\mathrm{bin}}$ as a function of radius at various times. Colorbar on the right indicates the time that each line corresponds to, which progresses as follows: dark brown to light orange (Epoch 1), dark blue to light green (Epoch 2), dark red to light purple (Epoch 3). The black dashed lines are the average of DiffRUN over Epoch 2 ($125 t_{\mathrm{bin}}<t-t_0<300t_{\rm bin}$). The black dotted lines are the average of LegacyRUN over the last $175t_{\rm bin}$($-54t_{\rm bin} < t-t_0 < 121t_{\rm bin}$). From the top to bottom: the ratio of shell-averaged thermal pressure to shell-averaged density, cooling rate per unit radius, {and cooling time}. (Left column) LegacyRUN, (right column) DiffRUN. The offset in $\shellint{\mathcal{Q}}$ at $t-t_0=0$ between DiffRUN and LegacyRUN reflects the internal energy doubling at the transition from LegacyRUN to DiffRUN (Sec.~\ref{sec: Simulation Details}).}
    \label{fig: thermal time smoothed}
\end{figure*}

{There are only two differences between %
LegacyRUN and DiffRUN, and both have to do with their treatment of thermodyamics.
DiffRUN assumes complete thermodynamic equilibrium in the sense that radiation contributes to the internal energy and pressure; and it substitutes a zeroth-order approximation for radiative cooling for %
{more crudely estimated} cooling rates (including none at all).}
Therefore, the changes to the surface density (see Sec. \ref{sec: Azimuthally-Averaged Surface Density}) and magnetic quantities (see Sec. \ref{sec: Evolution Magnetic}) must originate from this change to the system's thermal properties.
{In this section we  describe specifically how these properties changed.} 

Fig. \ref{fig: thermal time smoothed} displays the time evolution of: the thermal pressure \footnote{In this paper, thermal pressure $p_{\rm th}$ refers to the sum of gas pressure $p_{\rm gas}$ to the radiation pressure $p_{\rm rad}$.} per density $\shellavg{p_{\mathrm{th}}}/\shellavg{\rho}$ (i.e., to within a factor $\sim O(1)$, the square of the acoustic wave speed $c_s^2$); the cooling rate per unit radius $\shellint{\mathcal{Q}}$; and the cooling time $\shellint{u}/\shellint{\mathcal{Q}}$. They are shown as a series of time-smoothed line plots. %

Because it is in a quasi-steady state, LegacyRUN's thermal quantities, {especially its pressure/density ratio, change very little over its final %
$121t_{\rm bin}$.} Time smoothed profiles of $\shellavg{p_{\mathrm{th}}}/\shellavg{\rho}$ stay nearly constant.

Although the initial state is derived from the quasi-steady state of LegacyRUN, it is not identical because, as described in Sec.~\ref{sec: Simulation Details}, it was necessary to interpose a brief ``buffer'' run between the LegacyRUN snapshot that we used as a base and the true start of DiffRUN in order to avoid numerical instabilities triggered by an abrupt change in EOS.  The qualitative effect of the new thermodynamics model is to increase the cooling rate, which is why, in the range of radii with which we are particularly concerned, the initial sound speed-squared in DiffRUN is a factor $\sim 2$ smaller than in LegacyRUN, and the cooling rate in DiffRUN is larger by a factor $\sim 2$.  The initial cooling time changes much less because it depends principally on the vertical density structure, and the buffer run was not long enough (about $2t_{\rm bin}$) for this to change significantly. All of these contrasts diminish for $r \gtrsim 3.5a$.

Close examination of the curves in Fig.~\ref{fig: thermal time smoothed} reveals that the evolution of DiffRUN's thermal properties follows the same pattern as its magnetic properties: significant change during Epoch 1, a relatively inactivity during Epoch 2, and a return to more activity during Epoch 3.

{In the initial stage, when the magnetic stress is comparatively large, it does more work on the magnetized fluid, which leads to greater dissipation; as a result, first the ratio of thermal pressure to density rises, and then so does the cooling rate.  However, once diffusion cooling has been turned on, the ratio of thermal pressure to density at any given radius never exceeds the value found without cooling.  When the magnetic stress has been large for a long enough time to reduce the surface density, the cooling time, which is proportional to surface density, falls, inducing a decrease in the thermal pressure. }

{The events of this stage can also be viewed from the point of view of changes to thermal properties.  As already remarked, the thermal pressure per unit mass is always smaller than it was in LegacyRun, and at its lowest, smaller by almost a factor of 2.  The decrease in pressure leads to vertical compression of the gas, which amplifies the strength of the magnetic field's horizontal components, thereby increasing $p_{\rm mag}$.  Thus, increased magnetic pressure is a direct result of decreased thermal pressure, and the ratio between them, $p_{\rm mag}/p_{\rm th}$ rises from $\simeq 1/8$ in LegacyRun to $\simeq 1/2$ during the first stage of DiffRUN (compare the values shown in Fig.~\ref{fig: comparison of development} to those shown in Fig.~\ref{fig: thermal time smoothed}).  Thus, by increasing the cooling rate, our more physical treatment of disk thermodynamics leads to making the magnetic field dynamically important, which it had not been previously.  As such, it also resists further vertical compression.} 

{In the intermediate stage, the magnetic stress per unit mass is greater than in LegacyRUN, but not as large as it was in the initial stage.  During this period, the ratio of thermal pressure to density is only $\sim 1/3 - 1/2$ of what it was in the first part of the simulation because the cooling time is shorter.  With this fall in thermal pressure, $p_{\rm mag} \simeq p_{\rm th}$. The cooling rate is also lower in this stage than the first one, but the contrast is smaller than the contrast in pressure because the faster photon diffusion acts on a smaller photon energy density.}

{In the final stage, the magnetic stress is once again strong. Once again, there is greater dissipative heating. The cooling time decreases because the surface density declines, but only slightly because magnetic support prevents significant change in the scale height and the diminution of surface density is not very large. The net result of the increased heating is therefore a modest rise in the thermal pressure.}

\vskip 1cm

\subsection{Non-axisymmetric Disk Structures}
\label{sec: Non-axisymmetric Disk Structures}

Unlike in a single black hole accretion disk, the time-dependent quadrupolar potential of the black hole binary breaks the axisymmetry of the CBD. In this section we explain how the introduction of photon diffusion cooling alters non-axisymmetric structures.

\subsubsection{Elimination of the Lump}
\label{sec: Elimination of Lump}
\begin{figure*}[htbp]
    \centering
    \includegraphics[width=2\columnwidth]{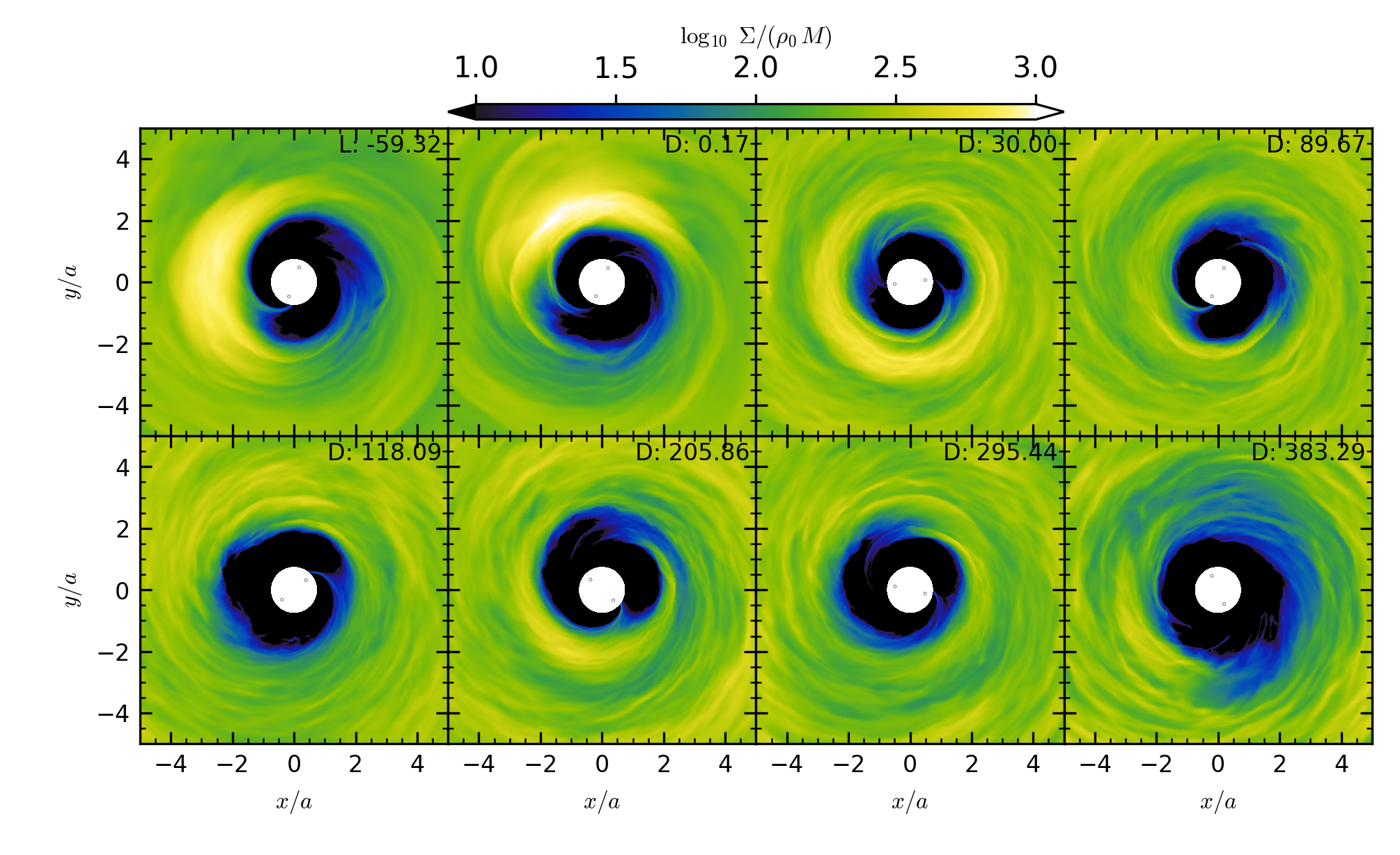} 
    \\[-11pt]
    \caption{Snapshots of $\Sigma$ at various times for DiffRUN and LegacyRUN. Color is in a logarithmic scale. 
    {In the upper-right corner of each panel, the letter indicates whether the snapshot is from LegacyRUN (L) or DiffRUN (D), while the the number denotes the value of $(t-t_0)/t_{\mathrm{bin}}$.}
    }
    \label{fig: surface density snapshots both}
    \centering
    \includegraphics[width=2\columnwidth]{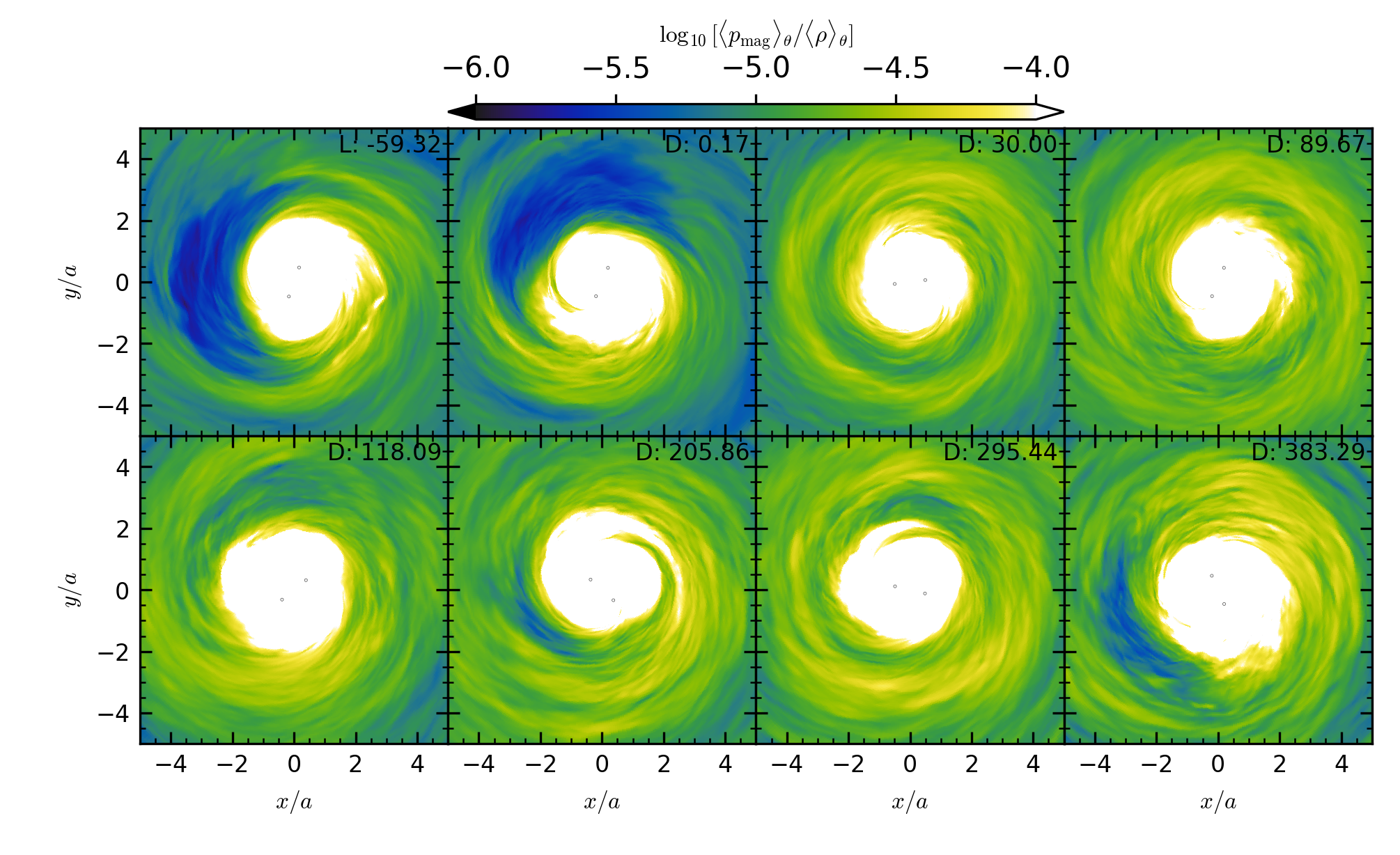}
    \\[-11pt]
    \caption{
{Like Fig.~\ref{fig: surface density snapshots both}, but showing $\shellavg{p_{\rm mag}}/\shellavg{\rho}$ instead.}
}
    \label{fig: pmag over density snapshots both}
\end{figure*}

Because the disks are thin ($H/r \lesssim 0.1$), %
{integrating over polar angle} approximates vertical integration well.
Fig.~\ref{fig: surface density snapshots both} shows the surface density $\Sigma(t,r,\phi)=\polarint{\rho}$ and Fig.~\ref{fig: pmag over density snapshots both} shows the ratio of vertically-averaged magnetic pressure to density $\polaravg{p_{\mathrm{mag}}}/\polaravg{\rho}$.

The most striking change over time that the images of surface density portray is a smoothing out of the very strong azimuthal concentration found in LegacyRUN: somewhere between $30t_{\rm bin}$ and $90t_{\rm bin}$ into DiffRUN, this concentration essentially disappears.
Similarly, the ratio $p_{\rm mag}/\rho$ has a deep {\it minimum} where $\Sigma$ has a {\it maximum}, and this minimum is filled in during the same time range that the maximum is erased.

{The surface density concentration is so} consistently reported in previous equal-mass CBD simulations {that it has been given a name: the ``lump" \citep{Shi2012}.}
The lump is distinguished from other density perturbations typical in an accretion disk, such as tidally induced density waves, by the fact that its pattern velocity matches the local Keplerian orbital velocity. Because the lump co-rotates with the disk, {it is a genuine physical object, not a pattern, and} its shape stays nearly steady over secular timescales. Lumps are also reported to be less magnetized than neighboring regions \citep{Noble2012, Noble2021, Tiwari2025EqualMass}, {just as shown in Figs.~\ref{fig: surface density snapshots both} and \ref{fig: pmag over density snapshots both}.} 

The lump in LegacyRUN has properties consistent with previous findings. During the last $\sim 200 t_{\rm bin}$ of the simulation, we observe the lump spanning the radial range $r\simeq 2a - 4a$, with $\polaravg{\rho}$ larger than the neighboring region by a factor of $\sim 3$ and $\polaravg{p_{\rm mag}}/{\polaravg{\rho}}$ smaller than the neighboring region by a factor of $\gtrsim 5$. The lump orbits at a frequency $\simeq 0.2 \Omega_{\rm bin}$, in %
close agreement with the orbital frequency at $r \simeq 3a$. The lump appears to source a strong $m=1$ density wave, which spirals outwards from its tail. These features can also be seen in the initial state of DiffRUN (see the third panel of Fig.~\ref{fig: surface density snapshots both} and the third panel of Fig.~\ref{fig: pmag over density snapshots both}).

{Nonetheless, despite possessing a large-amplitude lump strongly resembling the one seen in many previous simulations with unphysical cooling rates, in DiffRUN the lump is eradicated within several tens of $t_{\rm bin}$ within Epoch~1.}

{Elimination of the lump follows directly from the elimination of the surface density maximum} because most of the mass contained in this peak is found in the lump. 
As a result, during Epochs 2 and 3, DiffRUN shows no dominant {azimuthal} density concentration that persists across hundreds of $t_{\rm bin}$, in sharp contrast to LegacyRUN and to most previous MHD simulations of equal-mass CBD. Some snapshots display transient azimuthal asymmetry of $\Sigma$ near the inner edge, but these are irregularly shaped, short lived (a few dynamical timescales), and deform dynamically. The region with larger $\Sigma$ tends to have smaller $\polaravg{p_{\rm mag}}/\polaravg{\rho}$, but the contrast to the neighboring region {is at most} a factor $\sim 2$, {considerably smaller} than in the lump of LegacyRUN.

\subsubsection{Eccentricity of the Time-Averaged Disk}
\label{sec: Eccentricity of Time-Smoothed Disk}
\begin{figure}[htbp]
    \includegraphics[width=1.\columnwidth]{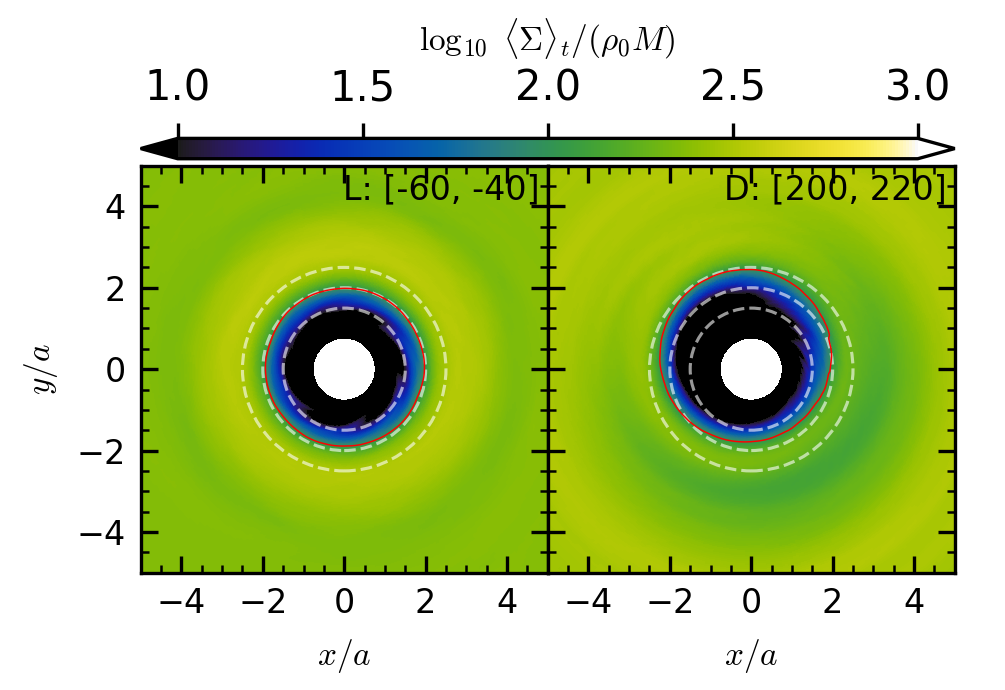}
    \caption{Time-averaged $\Sigma$ for DiffRUN and LegacyRUN. Color is on a logarithmic scale. White dashed circles are added to indicate $r= 1.5a$, $2.0a$, and $2.5a$. Red solid contours denote where $\timeavg{\Sigma} = 100\rho_0M$. (Left) LegacyRUN averaged from $t-t_0 = -60t_{\rm bin}$ to $-40 t_{\rm bin}$. (Right) DiffRUN averaged from $t-t_0 = 200t_{\rm bin}$ to $220 t_{\rm bin}$.}
    \label{fig: time averaged sigma}
\end{figure}

Does the lack of a lump make the radiation-pressure-dominated disk more axisymmetric? Not necessarily, because a different sort of azimuthal asymmetry develops in DiffRUN. {To see it clearly requires averaging the surface density over a few tens of}
dynamical timescales. This window is long enough to eliminate most of the density fluctuations that move at speeds comparable to Keplerian velocities, such as accretion streams, density waves, and the lump. {Measuring $\Sigma$ in this way (see Fig.~\ref{fig: time averaged sigma}) reveals that}  the disk's inner edge is {\it more} eccentric in DiffRUN than in LegacyRUN.
 LegacyRUN, despite its conspicuous azimuthal asymmetry in snapshots (see Fig. \ref{fig: surface density snapshots both}), becomes almost axisymmetric once the lump's orbital motion is integrated out. In particular, the contour of $\timeavg{\Sigma} = 100\rho_0M$, which approximates the edge, is nearly a circle of radius $2a$. In sharp contrast, DiffRUN's time-smoothed surface density distribution is substantially more azimuthally asymmetric. The inner edge of the disk is {distinctly} non-circular, with an eccentricity of $\sim 0.2$.
 {Put another way,} the radial coordinate of the $\timeavg{\Sigma} = 100\rho_0M$ contour  ranges from $r\simeq 1.75a$ to $r \simeq 2.5a$.
\begin{figure}[htbp]
    \centering
    \includegraphics[width=1\columnwidth]{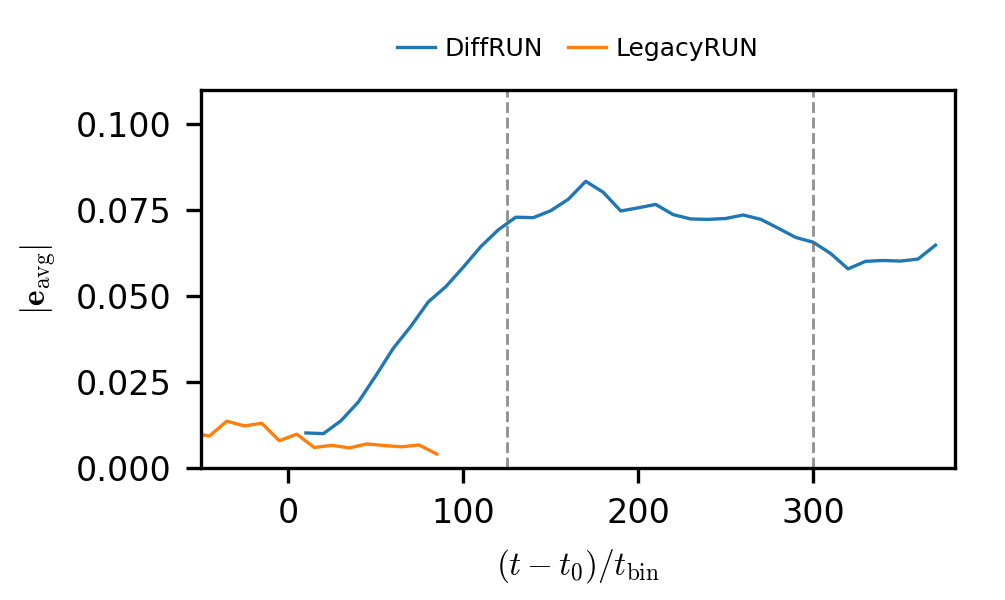}
    \caption{Magnitude of the box-car time-smoothed eccentricity vector $|\mathbf{e}_{\mathrm{avg}}|$ as a function of time. Vertical lines mark epoch boundaries.}
    \label{fig: longer term eccentricity vector}
\end{figure}

To study {more closely} the non-circularity of the disk orbits near the inner edge, we define an azimuthally averaged eccentricity vector $\mathbf{e}={\rm e}_x \hat {x} + {\rm e}_y \hat {y}$ as follows:
\begin{equation}
    {\rm e}_x(t) + i {\rm e}_y(t) = \frac{\int^{3a}_{1.5a} \dd r\shellint{i\rho (\sqrt{g_{rr}}v^r)e^{i\phi}}}{\int^{3a}_{1.5a} \dd r\shellint{\rho (\sqrt{g_{\phi\phi}}v^{\phi})}}.
    \label{eq: eccentricity}
\end{equation}
Here, the $x$-axis ($y$-axis) aligns with $\phi=0$ ($\phi=\pi/2$).
{In other words, the eccentricity vector is the ratio of the $m=1$ component of the radial momentum's Fourier transform to the azimuthal average of the azimuthal momentum in the inner portion of the CBD.}
{ If the orbit is eccentric and time-steady,} the eccentricity vector is parallel to the apsidal axis.

However, the lump deflects the orbital trajectories and thus perturbs the orbits of neighboring fluid elements on roughly the dynamical timescale. We suppress the contribution from this perturbation by defining a time-smoothed eccentricity vector, averaged over $20t_{\rm bin}$:
\begin{equation}
	\mathbf{e}_{\mathrm{avg}} \equiv \timeavg{{\rm e}_x} \hat{x} + \timeavg{{\rm e}_y} \hat{y}~.
\end{equation} 

We display the evolution of the
time-smoothed magnitude of the eccentricity in Fig. \ref{fig: longer term eccentricity vector}.
In the quasi-steady state of LegacyRUN, $|\mathbf{e}_{\rm avg}|$ is roughly $\sim 0.01$. Once we introduce a physical EOS and cooling scheme, $|\mathbf{e}_{\rm avg}|$ grows rapidly, increasing by a factor of $\sim 6$ by the end of Epoch 1. During Epochs 2 and 3, $|\mathbf{e}_{\rm avg}|$ 
{changes much less,}
staying within the range $0.05 - 0.08$ over hundreds of $t_{\rm bin}$.

{By constructing a toy-model,} {we speculate that the growth in eccentricity results from the decline in surface density near the CBD's inner edge. Following the methods of \cite{Lubow1994} and \cite{Shi2012}, we model the inner disk as a ring of mass $M_r$} {and argue, as did \cite{Shi2012}, that the eccentricity of an CBD is excited by stream impacts, and that the rate of excitation is $\propto \dot{M}_s/M_r$, where $\dot{M}_s$ is the mass flux of the stream.  We further suppose that eccentricity decays because fluid elements on eccentric orbits undergo adiabatic compression as they approach pericenter.  This compression coverts
orbital energy to internal energy, which can be lost to radiation.}
{Combining these elements leads to an equation for the rate of change of inner-edge eccentricity:}
\begin{equation}
    \frac{\dd e }{\dd t} = k\frac{\dot{M}_s}{M_r} - g(e(t)) ~,
\end{equation}
where $k$ is a positive constant and $g(x)$ is a positive, monotonically increasing function. The equilibrium eccentricity is then determined by 
\begin{equation}
    e_{\rm eq} = g^{-1}\left(k\frac{\dot{M}_s}{M_r}\right)~.
\end{equation}
Because the mass inflow rate in the two simulations is similar, $\dot{M}_s$ should also be similar. During Epoch 1, $M_r$ in DiffRUN diminishes from the quasi-equilibrium value of LegacyRUN, raising $e_{\rm eq}$. 
As $M_r$ levels out during Epoch 2, %
$e$ approaches $e_{\rm eq}$.

\subsubsection{Temporal Modulations}
\label{sec: Temporal Modulations}
\begin{figure*}[htbp]
    \centering
    \includegraphics[width=1\linewidth]{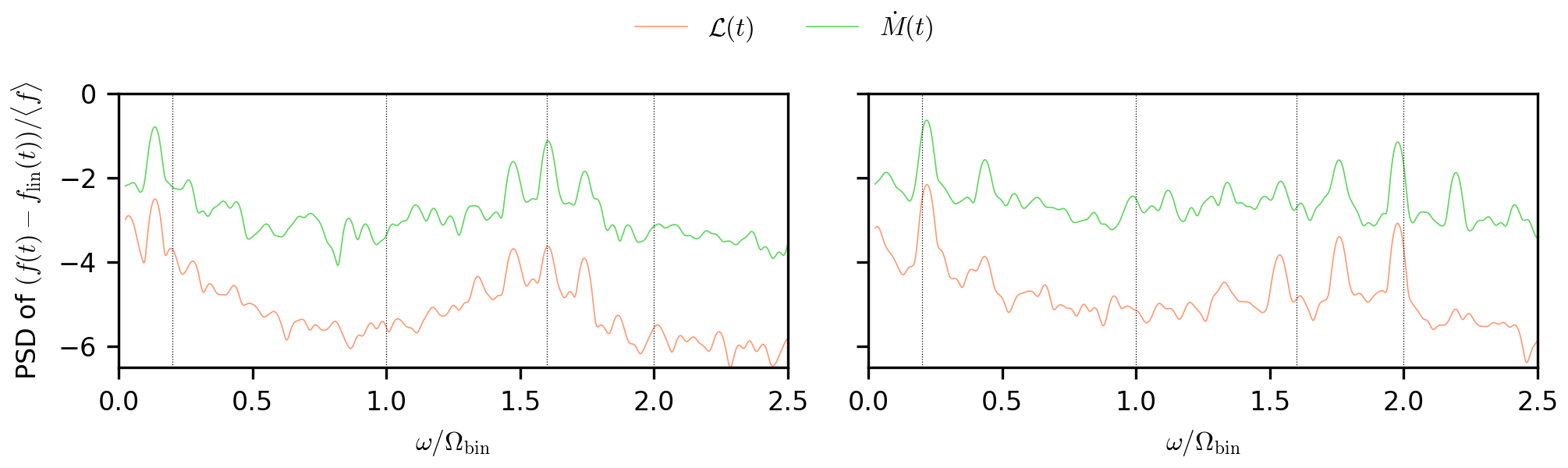}
    \caption{Line plots on a logarithmic scale (with base 10) of the temporal spectra of total luminosity $\mathcal{L}$, and inflow rate $\dot{M}$. (Left) LegacyRUN, (right) DiffRUN.
    Dotted vertical lines, {from left to right,} are at $\omega/\Omega_{\rm bin} = 0.2$ (lump frequency), $1.0$ (binary frequency), $1.6$ (twice the beat frequency), and $2.0$ (twice the binary frequency). 
    }
    \label{fig: temporal modulation}
\end{figure*}
Non-axisymmetric structures in CBDs break the continuous time-shift symmetry in the co-rotating frame of the binary. Instead, the system acquires quasi-periodicity, which may produce EM signal modulations. As previously remarked, the introduction of physical EOS and photon diffusion cooling changes the non-axisymmetric structures of the disk. In this section, we present how the temporal modulations of DiffRUN differ from those reported in previous studies.

To identify possible EM signal modulations, Fig.~\ref{fig: temporal modulation} displays the power spectral densities (PSDs) of the mass inflow rate $\dot{M}$ and total luminosity $\mathcal{L}$. %
Following \cite{Noble2012}, we define the total luminosity as
\begin{equation}
\label{eq: total lum}
    \mathcal{L}(t) =
    \int \dd r~ \dd \theta ~\dd \phi~\sqrt{-g}~\mathcal{Q} u_t.
\end{equation}
{Each PSD is obtained by subtracting a linear fit $f_{\rm fit}(t)$ from the raw data $f(t)$, normalizing by the time average $\langle f \rangle_t$, and then applying Welch's method with a boxcar window function of $15t_{\rm bin}$ length and $50\%$ overlap.  The time intervals for the transforms are}
Epoch 2 for DiffRUN and the last $175 t_{\rm bin}$ for LegacyRUN.  %

Previous MHD simulations of CBDs \citep{Shi2012, Noble2012, Noble2021, Tiwari2025EqualMass} report that the lump modulates EM signals at two characteristic frequencies: twice the beat frequency {between the binary orbital frequency and the lump's orbital frequency i.e., $2(\Omega_{\rm bin}-\Omega_{\rm lump})$ {when $q=1$}; and a low frequency near $\Omega_{\rm lump} \simeq 0.2 \Omega_{\rm bin}$.} LegacyRUN's PSD is consistent with this picture. 
{The PSDs of both} $\dot M$ and $\mathcal{L}$ peak at $\omega \simeq 1.6\Omega_{\rm bin} \simeq 2(\Omega_{\rm bin}-\Omega_{\rm lump})$.
In addition, a low frequency modulation of $\dot M$ and $\mathcal{L}$ appears at $\omega \simeq 0.14\Omega_{\rm bin}$, in the neighborhood of the lump frequency. %
From this perspective, it is expected that both characteristic frequencies disappear with the eradication of the lump. 

However, {as shown in Fig.~\ref{fig: temporal modulation},} even without the lump, DiffRUN continues to modulate $\dot{M}$ and $\mathcal{L}$ {with fractional amplitudes and contrasts with the continuum power similar to those in LegacyRUN}. Instead, the difference between these two can be found from the frequencies of these modulations, as we explain below.

{In both LegacyRUN and DiffRUN, the modulation frequencies can be described as combinations of two fundamental frequencies, which we call $\omega_1$ and $\omega_2$.  In LegacyRUN, $\omega_1 \simeq 1.6 \Omega_{\rm bin}$ and $\omega_2 \simeq 0.14\Omega_{\rm bin}$; in DiffRUN, these change to $\omega_1 \simeq 2\Omega_{\rm bin}$ and $\omega_{2} \simeq 0.21\Omega_{\rm bin}$}\footnote{One may find it perplexing that the lower-frequency modulation occurs almost at the lump frequency, although DiffRUN lacks the lump. This is because both this modulation and the lump are governed by the dynamical timescale of the inner {edge of the CBD}.} 
At low frequencies, $\dot{M}$ and $\mathcal{L}$ are modulated most strongly at $\omega_2$ in both runs, and in DiffRUN, to a lesser degree, at $2\omega_2$. {At higher frequencies, {modulations of} $\dot{M}$ in both cases and $\mathcal{L}$ in LegacyRUN peak at $\omega_1$ and $\omega_1 \pm \omega_2$, whereas $\mathcal{L}$ in DiffRUN is modulated at $\omega_1$, $\omega_1-\omega_2$, and $\omega_1 - 2\omega_2$.} {We note that signal patterns at $\omega_1$ and $\omega_1 \pm \omega_2$ are characteristic of a carrier signal of $\omega_1$ whose amplitude is modulated at a message frequency of $\omega_2$.}

It may appear strange %
that in DiffRUN the modulation frequencies do not match exactly between $\dot M$ and $\mathcal{L}$, but {\it a priori} $\dot{M}$ and $\mathcal{L}$ are two different non-linear responses to the underlying physical processes. %
$\dot M$ is directly modulated by the %
dynamical processes that generate the accretion streams; modulation of $\mathcal{L}$ is ultimately a thermal {\it response} to periodic stream impacts. In LegacyRUN, this response is tied to the change of local entropy, which is an immediate consequence of stream impacts. However, in DiffRUN, the thermal response is a function of both a quantity closely related to the local entropy, the radiation energy density, and two global quantities, the optical depth and spatial length of the optimal path from a cell to the disk surface.
The peak in both the ${\cal L}$ and $\dot M$ power spectra for DiffRUN at $\simeq 0.4 \Omega_{\rm bin}$ is likely the first harmonic of $\omega_2$, and is due to these nonlinearities.  
The peak at $\omega_1 - 2\omega_2$ in the power spectrum of ${\cal L}$ is then a beat frequency between this harmonic and $\omega_1$.

{The origin of the two fundamental frequencies may be seen by examining the binary's gravitational potential.  It can be expanded in terms of multipoles as:}%
\begin{equation}
    \Phi = \Phi^{(1)}+\Phi^{(2)}_{\rm{avg}} + \widetilde{\Phi}^{(2)} + \dots~.
\end{equation}
where $\Phi^{(1)}$ is the monopole term, $\Phi^{(2)}_{\rm{avg}}$ is the time-averaged quadrupole, and $\widetilde{\Phi}^{(2)}$ is the time-dependent quadrupole. Our arguments rely on the {\it geometry} and {\it time-dependence} of these two quadrupolar terms. $\Phi^{(2)}_{\rm{avg}}$ is axisymmetric \citep{Zilhao2015}, whereas $\widetilde{\Phi}^{(2)}$ features two lobes aligned with the binary separation axis, and therefore rotates at $\Omega_{\rm bin}$. Both decrease steeply ($\propto r^{-3}$) with radius, so a fluid element in an eccentric orbit interacts with them most strongly at the periastron.

Although DiffRUN has no lump, azimuthal perturbations of surface density can survive for a few dynamical timescales.  {Any} such perturbation passes through the periastron {once every orbital} period, where it experiences the non-Keplerian effects of $\Phi_{\rm avg}^{(2)}$ most strongly. This explains the lower frequency modulation at $\omega \simeq 0.21\Omega_{\rm bin}$, which is {the orbital frequency for a semimajor axis $\simeq 3a$.}

On the other hand, the two lobes of $\widetilde{\Phi}^{(2)}$ sweep past the periastron approximately two times per binary orbit, because the apsidal axis is nearly stationary in the inertial frame (see Sec. \ref{sec: Eccentricity of Time-Smoothed Disk}). This produces the higher frequency modulation at $\omega \simeq 2\Omega_{\rm bin}$. More generally, the higher frequency modulations from both simulations follow the same kinematic relation,
\begin{equation}
    \omega = 2(\Omega_{\rm bin}-\Omega_{\rm pattern})~,
\end{equation}
which describes how frequently the lobes of $\widetilde{\Phi}^{(2)}$ interact with the disk's $m=1$ structure traveling at $\Omega_{\rm pattern}$. In LegacyRUN, this structure is the lump, hence $\Omega_{\rm pattern}=\Omega_{\rm lump}$ and $\omega \simeq 1.6 \Omega_{\rm bin}$. In DiffRUN, this structure is the eccentric disk, whose apsidal axis is stable over several dynamical timescales, hence $\Omega_{\rm pattern} \ll \Omega_{\rm bin}$ and $\omega \simeq 2 \Omega_{\rm bin}$ {(see \cite{Tiwari2025EqualMass} for a related point of view)}. {It remains to be understood why $\Omega_\mathrm{pattern} = \Omega_{\rm lump} \ne \omega_2$ for the LegacyRUN, a question we leave for future investigation.}

\subsection{{Vertical Structure}}
\label{sec: Vertical Structure}

\begin{figure*}[htb]
    \centering
    \includegraphics[width=2\columnwidth]{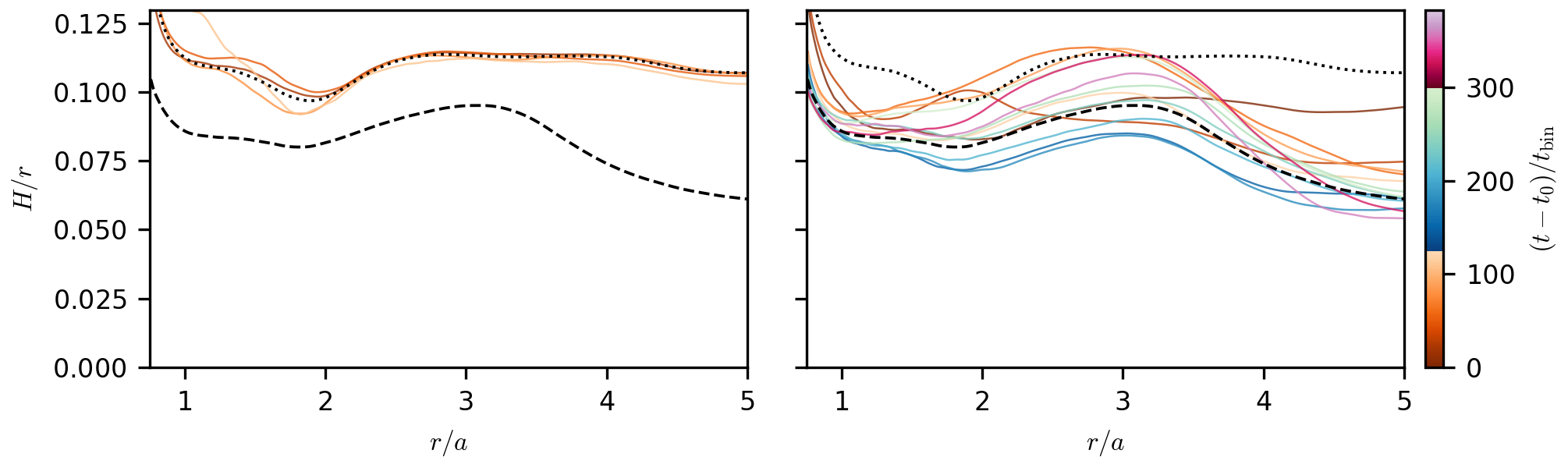}
    \caption{Aspect ratio of the disk, box-car smoothed
    over $30t_{\rm bin}$ as a function of radius at various times. Colorbar on the right indicates the time {range for each line:} %
    dark brown to light orange (Epoch 1), dark blue to light green (Epoch 2), dark red to light purple (Epoch 3). The black dashed lines are the averages of DiffRUN over Epoch2 ($125 t_{\mathrm{bin}}<t-t_0<300t_{\rm bin}$). The black dotted lines are the average of LegacyRUN over the last $175t_{\rm bin}$ ($-54t_{\rm bin} < t-t_0 < 121t_{\rm bin}$). (Left) LegacyRUN, (right) DiffRUN.}
    \label{fig: aspect ratio}
\end{figure*}

The introduction of diffusion cooling scheme and physical EOS alters how the energies are distributed among radiation, gas, and magnetic fields, resulting in a substantial shift in hydrostatic balances. In this section, we explain how the vertical structure of the disk changes and what supports the new structure.

For both DiffRUN and LegacyRUN, Fig.~\ref{fig: aspect ratio} displays the time evolution of $H/r$ as a series of time-smoothed line plots. DiffRUN inherits from LegacyRun an aspect ratio that increases slightly from $r \simeq 2a$ to $r \simeq 2.5a$ and is nearly constant at larger radii.
Outside $r \simeq 3a$, its subsequent evolution is an almost monotonic decrease, falling from $\simeq 0.11$ to $\simeq 0.06$ during the first $\simeq 150 t_{\rm bin}$.  Thereafter, it rises and falls with $0.06 \lesssim H/r \lesssim 0.08$.  Inside $r \simeq 3a$, the time-variation is more irregular, but converges toward $H/r \simeq 0.08$ at late times.  The initial diminution results from photon diffusion weakening radiation presure support; the ups and downs reflect fluctuations in the total pressure as magnetic energy density ($p_{\rm mag} = E_{\rm mag}$) is traded back and forth with thermal, mostly radiation, energy density ($p_{\rm rad} = E_{\rm rad}/3$).

\begin{figure*}[htbp]
    \subfloat{
    \includegraphics[width=1\columnwidth]{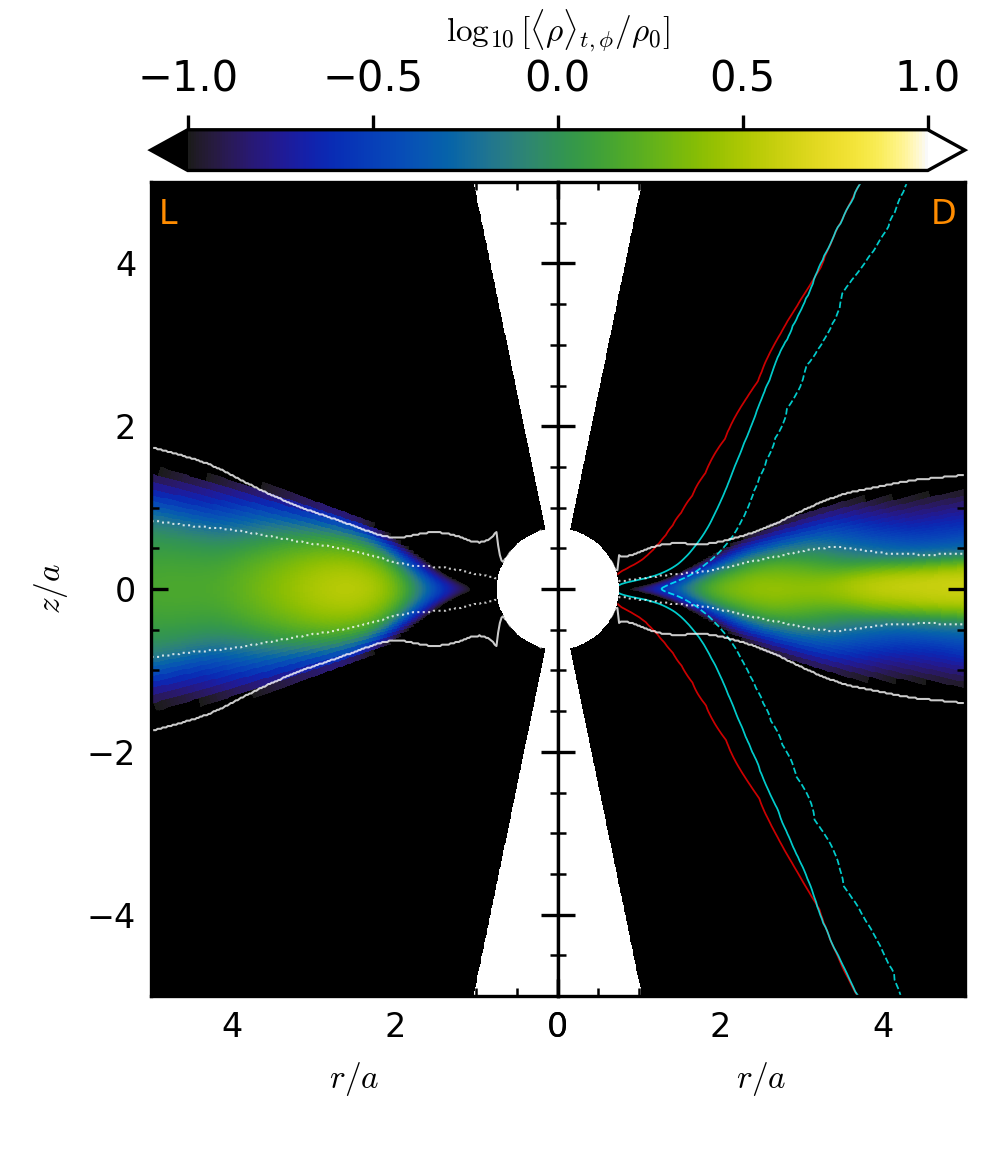}
    }
    \subfloat{
    \includegraphics[width=1\columnwidth]{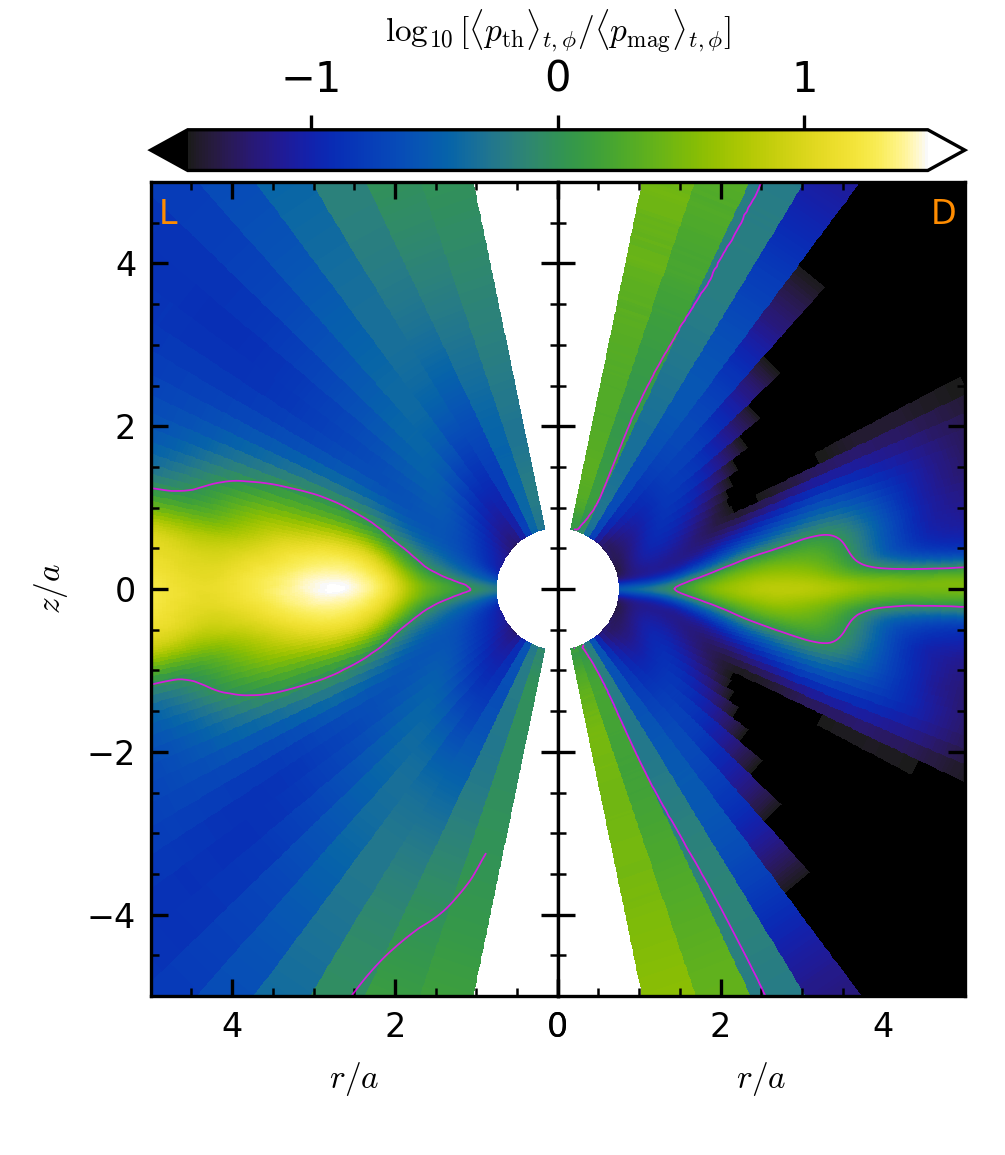}
    }
    \caption{Time and azimuthally averaged $\rho$ and $p_{\rm th}/p_{\rm mag}$ {(``plasma $\beta$")} on logarithmic colorscales. {(Left) $\timeazimavg{\rho}$, (right) $\timeazimavg{p_{\rm th}}/\timeazimavg{p_{\rm mag}}$.}  Each half panel is labeled with ``L" for LegacyRUN and ``D" for DiffRUN. LegacyRUN is averaged over the last $175t_{\rm bin}$ ($-54t_{\rm bin} < t-t_0 < 121t_{\rm bin}$), DiffRUN is averaged over Epoch 2 ($125 t_{\mathrm{bin}}<t-t_0<300t_{\rm bin}$). 
    {Several curves showing significant locations are also shown.} {The left panel (density): the surfaces enclosing (97\%, 80\%) of the vertically-integrated mass are shown by (solid, dashed) white curves for both simulations; in the DiffRUN half-panel, the surfaces where the time- and azimuthally-averaged $\timeazimavg{p_{\rm rad}}/\timeazimavg{p_{\rm gas}} = (1, 10)$ are shown by (solid, dashed) cyan curves, and the mass-weighted time- and azimuthally-averaged photosphere $\timeazimavg{\tau \rho}/\timeazimavg{\rho} = 1$ is shown by a red curve.  In the right panel (showing the plasma $\beta$), the surfaces where $\beta=1$ are shown by magenta curves.}}
    \label{fig: average density vertical}
\end{figure*}

\begin{figure*}[htbp]
    \centering
    \subfloat{
    \includegraphics[width=2\columnwidth]{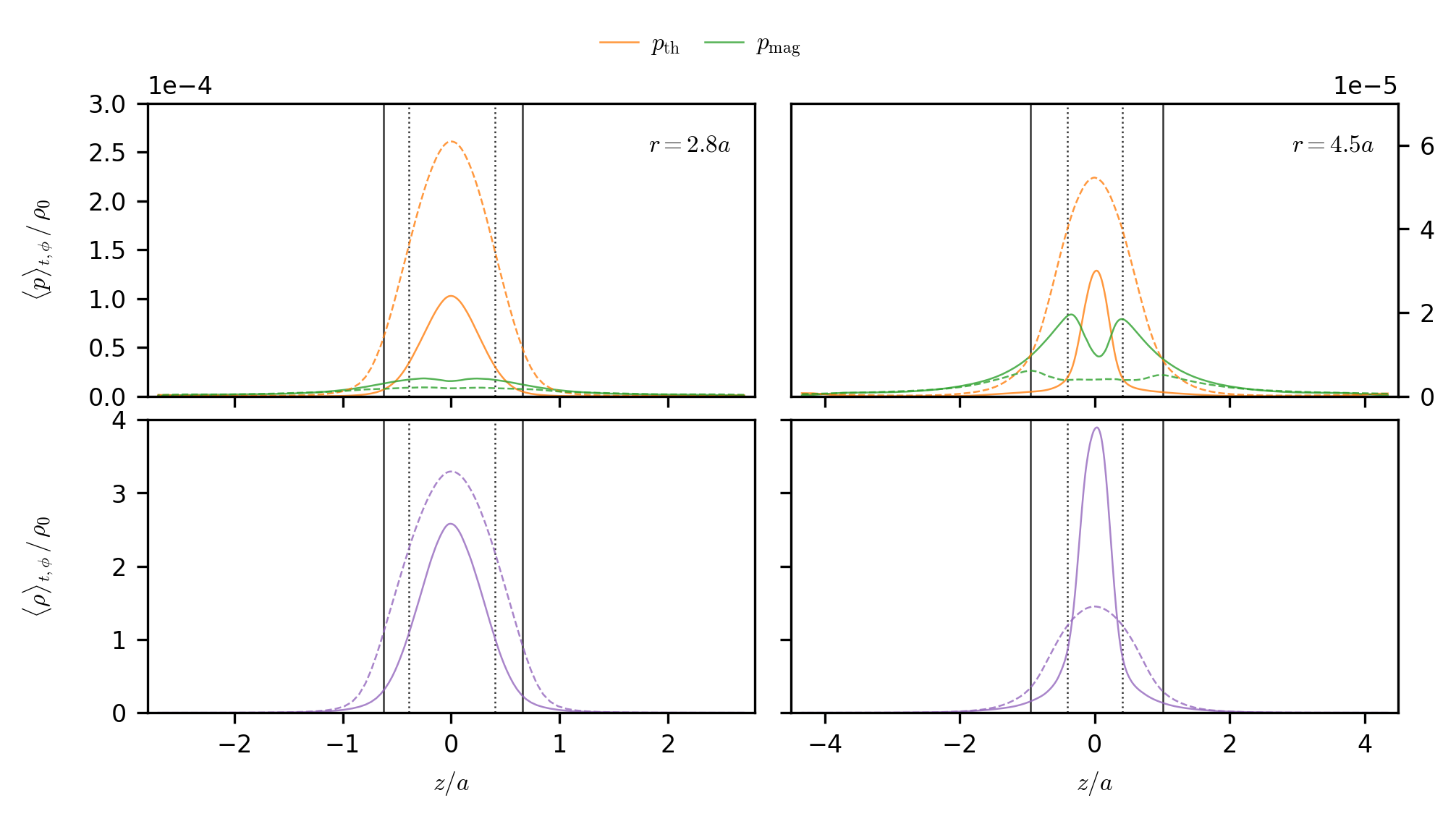}
    }
    \caption{Time and azimuthally averaged pressure and density as a function of the distance from the midplane. (Top) $\timeazimavg{p_{\rm th}}$, $\timeazimavg{p_{\rm mag}}$ on a linear scale, (Bottom) $\timeazimavg \rho$, $\rho_{\rm ad}$ on a linear scale. We plot DiffRUN and LegacyRUN on the same panel for the comparison. (Solid) DiffRUN, (dashed) LegacyRUN. Note that $p_{\rm th} \simeq p_{\rm rad}$, since $p_{\rm gas} \ll p_{\rm rad}$ in DiffRUN and $p_{\rm th} = p_{\rm gas}$, by construction, in LegacyRUN. 
    Vertical lines indicate the envelopes of the vertically integrated mass in DiffRUN: solid - $97 \%$ envelop, dashed - $80 \%$ envelop. Left column is at $r=2.8a$ and the right column is at $r=4.5a$. }%
    \label{fig: vertical support}
\end{figure*}

More detailed information on the vertical distribution of mass and pressure can be seen in Figs. \ref{fig: average density vertical} and \ref{fig: vertical support}. The former plots time and azimuthally averaged density $\rho$ and the ratio of thermal pressure to the magnetic pressure $p_{\rm th}/p_{\rm mag}$.  The latter provides more quantitative information on the vertical structure of the disk, as it displays time and azimuth averaged density, gas pressure, radiation pressure, and magnetic pressure at fixed radii $r=2.8a$ and $r=4.5a$ in line plots. Each radius is chosen to represent a typical vertical slice of the inner disk ($2a<r<3a$) and the outer disk ($4a<r<5a$). 

The %
pressure distributions in DiffRUN are substantially different from those in LegacyRUN.  As shown in Figs.~\ref{fig: average density vertical} and \ref{fig: vertical support}, the radiation pressure is much larger than the gas pressure in essentially the entire disk, whereas there is no radiation pressure at all in LegacyRUN.  In addition, although the typical plasma ``$\beta$''\footnote{In our paper, plasma $\beta$ the ratio of {\it thermal} pressure ($p_{\rm th} = p_{\rm gas} + p_{\rm rad}$) to the magnetic pressure ($p_{\rm mag}$).} is 
$\sim O(10-30)$ in LegacyRUN, it decreases to $\sim O(1 - 10)$ with diffusion cooling and  a LTE %
EOS. The magnetic field thus becomes dynamically relevant throughout the entire disk body.

Their specific effect on equilibrium disk structure, however, depends on the location, as illustrated in Fig. \ref{fig: vertical support}. At $r=2.8a$, $p_{\rm mag}$ has little direct contribution to time-averaged local hydrostatic equilibrium because it is evenly distributed vertically. However, plasma $\beta$ is small enough ($2 - 5$) that magnetic fields can, in principle, influence global structures of the disk. At $r=4.5a$, magnetic fields are even more dynamically important, directly participating to the local hydrostatic equilibrium.  In the upper layer ($|z| \gtrsim 0.5a$), plasma $\beta$ is generally $<0.3$, and $p_{\rm mag}$ gradient provides an almost entire support against gravity. In the core ($|z| \lesssim 0.5a$), $p_{\rm mag}$ gradient still is comparable to $p_{\rm th}$ gradient, but with an opposite sign. Here, $p_{\rm th}$ provides vertical support against gravity {\it and} $p_{\rm mag}$.

{Figs. \ref{fig: average density vertical} and \ref{fig: vertical support} display how {DiffRUN redistributes mass vertically.}
The disk of LegacyRUN retains a similar vertical structure throughout the disk body: at both $r=2.8a$ and $r=4.5a$, the vertical density distribution resembles a typical thermally-supported thin disk, and the {heights within which 80\% and 97\% of the mass are found
are roughly proportional to the} scaleheight $H$. 
In sharp contrast, the vertical structure of DiffRUN 
{changes significantly at $r \simeq 3 - 4a$ }
(Fig.~\ref{fig: average density vertical}).
At $2a<r<3a$, 
{the enclosed mass contours have heights proportional to $r$}
whereas at $4a<r<5a$, 
{the 97\% enclosed mass contour}
flares  
{while the 80\% contour} %
plateaus. %
{In this way, the outer disk ($4-5a$)}
stratifies into a {thin dense core containing $\simeq 80\%$ of the total mass, and an extended atmosphere containing the remainder.}}

\section{Discussion}
\label{sec: discussion}

\subsection{{Elimination} of Density Concentrations}
\label{sec: Elimination of Density Concentrations}

One striking implication of our research is that surface density features {in many previous simulations} of equal mass CBDs {disappear when radiation dominates the disk pressure and a physical cooling rate is employed.} %
Once we introduce the diffusion cooling scheme and physical EOS, the magnetic stress becomes stronger as the compression of disk enhances magnetic field. Consequently, over a timescale $\sim \mathcal{O}(100t_{\rm bin})$, the surface density distribution smooths out significantly
{in both the radial and azimuthal directions, eliminating}
the azimuthally-averaged surface density maximum as well as the lump. Throughout the quasi-steady state (Epoch 2) and the subsequent secular evolution until the end of simulation (Epoch 3), the disk continues to lack both the surface density maximum and the lump.

{To understand how these features are removed, we focus on the CBD's angular momentum budget.}
The defining feature of a CBD is that the gravitational torque from its central binary acts as a source of angular momentum. Transportation of this angular momentum dictates how the matter is distributed near the inner edge.

A 1d model of a CBD (\cite{Pringle1991}) predicted that the gravitational torque will delay accretion to such an extent that {\it no} material can pass the disk's inner edge. In this model, when the outer disk continues to supply material, the material accumulates just outside of the inner edge, resulting in an indefinitely growing surface density maximum. 2d $\alpha$-hydro and MHD simulations break the azimuthal symmetry and {allow a pair of accretion streams to drain mass from the CBD}.  As a result, the surface density just outside the inner edge saturates at a finite value. Nevertheless, a surface density maximum persists in the quasi-steady state of LegacyRUN, like in many previous MHD studies \citep{Shi2012, Noble2012, Noble2021} as well as in many more 2D viscous hydrodynamics calculations \citep{Artymowicz1991, Macfadyen2008, Farris2014, Miranda2017, Franchini2021}.

Like the surface density maximum, a lump has been commonly observed in both 2D $\alpha$-model hydrodynamics and 3D MHD simulations. %
Both MHD \citep{Shi2012} and $\alpha$-model hydrodynamics \citep{DOrazio2013} studies proposed that 
{the lump is built by an orbital timing-coincidence in which} the torqued stream sourced from the lump returns to the disk {close to where the lump is then located,} thus %
enhancing its density.

{Although lump and surface density maximum growth in previous MHD has been attributed to a hydrodynamical mechanism, those calculations have shown signs that magnetic stresses can compete with the hydrodynamical effects.}
{In} a quasi-steady state, shell-integrated %
angular momentum conservation demands {that} the gravitational torque density and the divergence of the Reynolds and Maxwell stresses must sum to zero.
Thus, when magnetic stresses increase, the relative influence of the gravitational torque decreases.
{In addition, as shown by \citet{Noble2021} for the lump, stronger magnetic stresses diminish the lump amplitude.}

From this perspective, the effect of physical cooling may then be summarized as an enhancement of the magnetic stress {to the point that it becomes the {\it principal} element in angular momentum transport, balancing the gravitational torque exerted by the binary.  Because { in these circumstances} the divergence of the magnetic stress is roughly equal to the binary torque {when the system reaches equilibrium}, and the magnetic stress per mass increases sharply with loss of radiation pressure by cooling, a smaller amount of mass is required in order to produce the stress needed for equilibrium.   A smoothed-out surface density profile is an immediate consequence, stemming directly from the comparatively larger role played by magnetic forces.}

\subsection{Effects on Binary's Orbital Evolution}

{
The binary's orbital evolution is determined by a competition between two physical processes of comparable, but opposite effects: accretion streams carry energy and angular momentum {\it to} the binary; the gravitational torque (on net) removes energy and angular momentum {\it from} the binary. {Because the net exchange of energy and angular momentum is determined by the competition between these oppositely-directed flows, it is very sensitive to their precise magnitudes.} These processes are %
likely to be affected at the quantitative level by the physical changes that the introduction of realistic thermodynamic model causes to the inner disk, {potentially creating a qualitative change to the {\it net} rates of energy and angular momentum transfer.}
{For example,
a fluid element detaches from the CBD when the combination of Reynolds and Maxwell stresses reduces its angular momentum slightly. The cooling changes the distribution of these stresses, as well as how close the periastron is to the binary, thereby altering the trajectory at which the accretion stream is launched into the cavity. This, in turn, changes the amount of angular momentum and energy gravitational torque delivers to the streams. 
After being torqued by the binary, some fraction of the streams are deflected towards the disk and subsequently impact it. The amount of energy and angular momentum that stream impacts deliver to the disk is affected by the increase in Maxwell stresses and reduction in mass of the inner disk.
{To evaluate quantitatively the net change in binary orbital evolution due to these effects requires a separate calculation.}
}

\subsection{{EM Emission Modulation}}

Many previous studies of CBDs, both MHD- \citep{Noble2021, Gutierrez2022, Tiwari2025EqualMass} and hydrodynamic- \citep{Farris2015Inspiral, Cocchiararo2024, Cocchiararo2026, Betancourt2026} based, have explored how the lump can modulate EM signatures of an equal mass binary system.
After accounting for radiation pressure and physical cooling, the lump is eliminated, but {the quadrupolar potential and eccentricity in the inner edge of the CBD} modulate $\dot{M}$ and $\mathcal{L}$ at least as strongly as the lump does. 

{One notable implication is that, even without the lump, a modulation of the lightcurve at a frequency $\sim \Omega_{\rm bin}$ is a strong signature of a binary system. However, it may still be challenging to infer %
the exact binary frequency $\Omega_{\rm bin}$ from the modulation alone, as the numerical factor may vary {considerably depending on which mechanism excites it and the system parameters. For example, as shown in Sec.~\ref{sec: Temporal Modulations}, there can be periodic variation of the light from a CBD at a variety of combinations of what we call $\omega_1$ and $\omega_2$, all of them $\sim \Omega_{\rm bin}$, but ranging from $\simeq 0.1\Omega_{\rm bin}$ to $\simeq 2\Omega_{\rm bin}$.
Modulation at $2\Omega_{\rm bin}$ rather than $\Omega_{\rm bin}$ is the signature of an equal-mass binary \citep{Shi2012}.
Similarly,} \citet{Tiwari2025UnequalMass} report a modulation occurring at $\simeq 0.5\Omega_{\rm bin}$ for a binary with a mass ratio $q=10$.}

{The preceding considerations were most relevant to our predicted modulation of $\mathcal{L}$, the luminosity from the CBD itself.  {Identification of a particular modulation with the CBD might be clearer if the modulation appears in a spectral band dominated by the CBD, rather than the minidisks, e.g., at photon energies below a possible ``notch'' in the continuum spectrum \citep{Roedig2014, Farris2015Thermal, Tang2018}.}

Modulation of $\dot M$, on the other hand can lead to %
luminosity fluctuations it drives in the minidisks.  For these modulations to be observable, however, two conditions must be met. 
One is that the modulated radiation must emerge from 
locations the accreting matter reaches within a time $< t_{\rm bin}$ from when it arrives at the minidisk and having cooling times short compared to $t_{\rm bin}$. 
The second is that the luminosity modulated cannot be too small compared to the luminosity radiated in the same band from other portions of the system.  How much energy is available to radiate in the minidisks depends on the potential depth at these regions (and therefore the system separation and mass-ratio) and the efficiency of heat-generation (e.g, through shock geometry, etc.). }

\subsection{{Physical Cooling and Vertical Structures}}

{MHD simulations of CBDs with simplified (adiabatic or isothermal) EOS and unphysical (or no) cooling {are generally defined so as to} rely on a single %
parameter (e.g. target sound speed or target entropy)
{that defines the scale height.  Hydrostatic equilibrium in the context of the specific model then determines the vertical density distribution in terms of $z/H$.}
This limitation is exemplified by LegacyRUN, which develops a thermally-supported disk with a nearly constant $H/r\simeq 0.11$.}

{
In our simulation, inclusion of radiative cooling and radiation pressure {dramatically expands the accessible phase space of possible structures because the cooling rate depends on $\Sigma$ and $H$, both of which are dynamical variables.  The specific impact due to introducing this more physical description of disk thermodynamics is an initial}
loss of thermal support, but the vertical compression that follows simultaneously enhances magnetic support. {One immediate consequence is to}
confirm previous radiative MHD (RMHD) simulations showing \citep{Tiwari2025EqualMass, Tiwari2025UnequalMass, Sadowski2016, Jiang2019, Liska2022, Zhang2025} %
that CBDs (and disks around single black holes) with $p_{\rm rad}\gg p_{\rm gas}$ are feasible {\it because} magnetic fields stabilize the disk against thermal instability, contrary to the {$\alpha$}-model's prediction that radiation pressure dominated disks are thermally unstable \citep{Shakura1976}.
}

{{In addition, the phase space expansion permits $H$ to be location- and time-dependent, and $\rho(z/H)$ can take many different forms.}
The inner part of the CBD we studied is relatively similar to a typical thermally supported disk, including a ratio $H/r$ nearly independent of radius, but the outer disk stratifies into a thin, dense, and thermally supported core and an extended, sparse, and magnetically supported atmosphere. {In this outer part, $H$, rather than $H/r$, becomes nearly independent of radius, but because $\rho(z/H)$ takes on quite different forms close to the midplane and farther away from it, the utility of defining a single value of $H$ becomes questionable.}
}

\subsection{{Comparison to CBD simulations with cooling}}

In this section, we make comparison to previous CBD simulations which introduced cooling. We focus on the binaries with equal mass and circular orbits.

\subsubsection{2D $\alpha$-hydro models}

Other sorts of cooling prescriptions have been adopted for use in 2D hydrodynamics simulations with $\alpha$-model ``viscosity''. Some are based on an approximate treatment of photon diffusion in LTE \citep{Farris2015Inspiral, Cocchiararo2026}; others \citep{Wang2023} take the simpler approximation of a timescale proportional to the local orbital timescale (the ``$\beta$'' ansatz). Because 2D simulations are typically $\gtrsim 100\times$ computationally cheaper than MHD simulations, a wide range of configurations has been explored using a version of this method. In these simulations, however, the inner disk structure does not respond to cooling as it does in our simulation.

For example, \cite{Wang2023} find that an {\it increase} in cooling timescale weakens the lump. Because they fix the target temperature, an increase in the cooling timescale (equivalently $\beta$) produces a hotter and thicker disk, so the disappearance of the lump occurs under the opposite thermodynamic condition to what we observe.

Even $\alpha$-model studies with diffusion cooling find either the lump or the surface density maximum to be a persistent feature of a CBD. Both \cite{Farris2015Inspiral} and \cite{Cocchiararo2024} implement an approximate diffusion cooling while retaining an EOS in which the (vertically integrated) photon pressure is ignored. 
As in many previous $\alpha$-model studies without cooling, the lump remains a persistent feature. \cite{Cocchiararo2026} improve this method by adopting a physical EOS with photon pressure, and report that the lump is eradicated. Other properties of
the disk, however, evolve opposite to ours: the eccentricity decreases and the surface density maximum grows.

Regardless of the thermodynamic scheme, to our knowledge no previous $\alpha$-hydro study has reported that an increase in the cooling rate causes elimination of {\it both} the surface density maximum and the lump.

The discrepancy between $\alpha$-model studies and our simulation is attributable to how the stress is determined in each methodology. In the $\alpha$-model, the stress is %
{assumed} to be proportional to the vertically integrated thermal pressure, so a colder disk necessarily has a weaker stress and suppressed angular momentum transport. In MHD simulations, the stress originates from the magnetic field, which grows as cooling compresses the disk. As suggested by this study, this effect alone can significantly alter the properties of a CBD, although it is opposite to what the $\alpha$-model predicts by construction. {Because magnetic fields are ubiquitous in accretion disks around black holes,} it thus appears that the $\alpha$-model is not appropriate {to the inner regions of CBDs.%
} 

\subsubsection{Radiative MHD}
\label{sec: Radiative MHD}
Perhaps the {previous work} most relevant to our {results} is a Newtonian RMHD simulation by \cite{Tiwari2025EqualMass}. Like us, they studied a radiation pressure dominated disk around a SMBH binary. %
{The most significant} methodological difference {between their work and ours is} that their simulation used {multi-angle group} radiation transfer to calculate the local cooling rate.  {This method} is more accurate {than ours, but entails the disadvantage of a large} %
increase in computational cost. {Their specific parameters were also somewhat different from ours: both of the simulations study binary separation of $100r_g$, but the mass of our binary is larger by a factor of $5$,
and our surface density %
is larger by a factor of $\sim 10^2$, resulting in a greater accretion rate in Eddington units ($\dot M/\dot M_{\rm Edd} \simeq 20$ vs. $\dot M/\dot M_{\rm Edd} \simeq 0.15$).
In addition, our disk is thicker ($H/r \sim 0.1$ vs. $H/r \sim 0.03$ at $r=3a$) and less magnetized.

{The first $\sim 30 t_{\rm bin}$ of our simulation strongly resembles what they found in the $53 t_{\rm bin}$ duration of their simulation.
In both, physical radiative cooling significantly reduced the disk's scale height, compressing the magnetic field sufficiently to make it dynamically important; as the magnetic field strengthened,
the lump's density contrast against the background diminished and it was elongated azimuthally; $H/r$ acquires a significant radial dependence (decreases with radius).} 

{However, the methodological difference we mentioned previously led to Tiwari et~al. stopping their simulation at $\simeq 50 t_{\rm bin}$, a time much shorter than their inflow time $t_{\rm in} \equiv  9\pi a^2~\Sigma(3a) /\dot M \sim 10^3 t_{\rm bin}$ (in ours, $t_{\rm in} \simeq 300 t_{\rm bin}$).  Consequently, they were unable to follow any significant rearrangement of the radial surface density profile.  By contrast, our more approximate, but computationally cheaper, physical cooling rate permitted us to run for $\simeq 380 t_{\rm bin}$, long enough to see surface density evolution.}

\section{Conclusions}
\label{sec:conclusions}

{In this paper, we have presented a radiative 3D MHD simulation of an equal mass circumbinary disk.  {In sharp contrast with previous efforts, it employs a physically-motivated cooling rate that is sensitive to the 3D density and temperature structure of the disk.}
This method captures essential thermodynamical effects that unphysical cooling schemes neglect at a substantially lower computational cost than the radiative transfer method. We began our simulation from a quasi-steady state of a target-entropy cooling MHD simulation, and evolved it until the disk reached a quasi-steady state, which it maintains for about $175$ binary orbital periods.}

{{The conditions of the target-entropy simulation imply a large ratio of radiation to gas pressure if the gas is in LTE.} The immediate effect of introducing radiation pressure and physical cooling is that the disk loses thermal support and compresses vertically, amplifying the magnetic field, as previously found in 3D MHD simulations employing full radiation transfer
\citep[e.g.][]{Tiwari2025EqualMass, Zhang2025}. }

{{The main new findings of our paper, {likely applicable to any CBD in which radiation pressure is dynamically important,} arise from the secular evolution over at least an inflow time, a time long enough for the radial mass distribution to change.}
We summarize them as follows:
\begin{itemize}
    \item {Most importantly,} near the CBD's inner edge, the magnetic stress per unit mass increases significantly, so that these stresses transport angular momentum more rapidly. Both the lump and the azimuthally-averaged surface density maximum are smoothed out to an extent that they are completely eradicated.  {This finding demonstrates that a more physical treatment of disk thermodynamics has a qualitative effect on its structure.}
    \item The time-averaged inner edge becomes more eccentric because with less mass in the region the {orbits of} inner disk {matter are}
    more susceptible to %
    stream impacts.
    \item Eccentricity of the disk generates modulations of accretion rate and disk luminosity at $2 \times$ and $0.2 \times$ the binary frequency, in place of the lump beat frequency (at $1.6 \times$ the binary frequency) modulation.
    {These effects are potentially observable.}
    \item As remarked in Sec.~\ref{sec:introduction}, the binary evolution depends sensitively on the dynamics of energy and angular momentum transportation from the binary to the disk. Introduction of {physical thermodynamics and the concomitant change in surface density profile very likely} affects binary evolution.
    \item The vertical density distribution depends strongly on the radius: the inner disk is {primarily thermally-supported (by radiation pressure)}; the outer disk stratifies into a thermally-supported core and a magnetically-supported atmosphere.  
\end{itemize}
}

\begin{acknowledgments}
\label{Sec:ack}
This work was partially supported by NASA TCAN grant 80NSSC24K0100. {A portion of this work took place at the Aspen Center for Physics, which is supported by National Science Foundation grant PHY-2210452. LLMs (Claude Opus 5 and Fable 5.1) were used to suggest language improvements. AI coding assistants (Github Copilot and Anthropic Claude code) were used in developing the simulation, data analysis, and visualization codes. Responsibility for the final manuscript lies entirely with the authors.} %
\end{acknowledgments}

\appendix

\section{Calculating temperature from optical depth and internal energy density}
\label{sec: EOS temperature appendix}
{
The gas temperature $T$, given the EOS Eq. \ref{eq: internal energy}, satisfies the quartic equation:
\begin{align}
    T^4+a_1T +a_0 &= 0~\\
    a_1 &\equiv \frac{3 \rho k_B}{2f_{\rm eff}(\tau) a_{\rm SB}\overline{m}}\\
    a_0 & \equiv -\frac{u}{f_{\rm eff}(\tau)a_{\rm SB}}
\end{align}
}
{The $\rho$ and $u$ floors, which are described in Sec.~\ref{sec: Simulation Details}, demand that $a_1>0$ and $a_0<0$; it follows that there is a unique real positive $T$. Ferrari's method gives an exact analytic expression for it:
\begin{align}
    \label{eq: quartic soln}
    T &= \frac{\sqrt{s}}{2}\Bigg(-1 + \sqrt{-1 + 2\sqrt{1+X_2}}\Bigg)~.
\end{align}
Here, we introduce auxiliary variables defined as:
\begin{align}
    X_1       &\equiv -\frac{27 a_1^{4}}{256 a_0^{3}}~,\\
    u_{\pm} &\equiv \frac{a_1^{2}}{2}\Bigg(1 \pm \sqrt{1+\frac{1}{X_1}}\Bigg)~,\\
    s       &\equiv u_{+}^{1/3} + u_{-}^{1/3}~,\\
    X_2       &\equiv -\frac{4a_0}{s^{2}}~.
\end{align}
}

{
Evaluated directly in the code, the analytic expression for $s$ loses accuracy at small $X_1$ and the analytic expression of $T$ loses accuracy at small $X_2$ because the expressions involve a subtraction between two close values. We therefore use asymptotic expressions in these limits:
\begin{align}
    T &\simeq \frac{\sqrt{s}}{4}X_2
      \left(1 - \frac{1}{2}X_2 \right) + \mathcal{O}(X_2^3)~,
      \qquad X_2 \ll 1~,\\
    s &\simeq \frac{\left(4a_1^{2}X_1\right)^{1/3}}{3}
      \left(1 - \frac{4}{27}X_1\right) + \mathcal{O}(X_1^{2})~,
      \qquad X_1 \ll 1~.
\end{align}
The expansions are used whenever the relevant quantity ($X_1$ or $X_2$) is $< 10^{-6}$; otherwise, 
Eqn.~\ref{eq: quartic soln} is used.}

\section{Analytic solution to the average cooling rate per cooling timestep}
\label{sec: Lambert-W function appendix}
{
If the diffusion time is regarded as fixed throughout the timestep, the evolution of internal energy is expressed by
\begin{equation}
\begin{split}
    u = AT^4 + BT, ~~\frac{du}{dt} = -CT^4, ~~ A, B, C>0,
\end{split}
\end{equation}
with
\begin{equation}
    A = f_{\rm eff}(\tau) a_{\rm SB}, \quad
    B = \frac{3\rho k_B}{2\overline{m}}, \quad
    C = \frac{\kappa \rho c}{\tau^{2}} f_{\rm eff}(\tau) a_{\rm SB}.
\end{equation}
We apply the separation of variables to find
\begin{equation}
    \left(\frac{4A}{T}+\frac{B}{T^{4}}\right)dT=-C \dd t~,
\end{equation}
which we integrate from $t_0$ to $t_0+\Delta t$ with $T_i\equiv T(t_0)$ and $T_f\equiv T(t_0+\Delta t)$:
\begin{equation}
\alpha Y+\beta \ln Y=\alpha + \delta~.
\end{equation}
Here, we have introduced four auxiliary variables:
\begin{equation}
   Y \equiv \left(\frac{T_i}{T_f}\right)^{3}, \quad
    \alpha\equiv \frac{B}{3T_i^{3}},\quad
    \beta\equiv \frac{4A}{3},\quad
    \delta \equiv C\Delta t . 
\end{equation}
This form implies that
\begin{equation}
    Y = \frac{\beta}{\alpha}W\left(\frac{\alpha}{\beta}e^{\frac{\alpha +\delta}{\beta}}\right),
\end{equation}
where the Lambert $W$ function is defined as
\begin{equation}
    W(x)e^{W(x)}=x.
\end{equation}
Letting
\begin{equation}
    c_1 \equiv \frac{\alpha}{\beta}
        = \frac{1}{4}\frac{u_{\text{gas}}(t_0)}{u_{\text{rad}}(t_0)}, ~~~
    c_2 \equiv \frac{\delta}{\beta}
        = \frac{3}{4}\frac{\kappa \rho c\, \Delta t}{\tau^{2}},
\end{equation}
we have
\begin{equation}
    Y = W(c_1 e^{c_1 + c_2})/c_1
\end{equation}
and
\begin{equation}
\begin{split}
    \Delta u &= \Bigg[u_{\text{gas}}\left(\frac{T_f}{T_i}-1\right)+u_{\text{rad}}\left(\frac{T_f^4}{T_i^4}-1\right)\Bigg]\\
    &= u(t_0)\Bigg[\frac{Y^{-1/3}(Y^{-1}+4c_1)}{4c_1 + 1}-1\Bigg]~.
\end{split}
\end{equation}
}

\vskip 1cm

\section{MRI resolution}
\label{sec: MRI resolution}

\begin{figure*}[htbp!]
    \centering
    \includegraphics[width=0.7\textwidth]{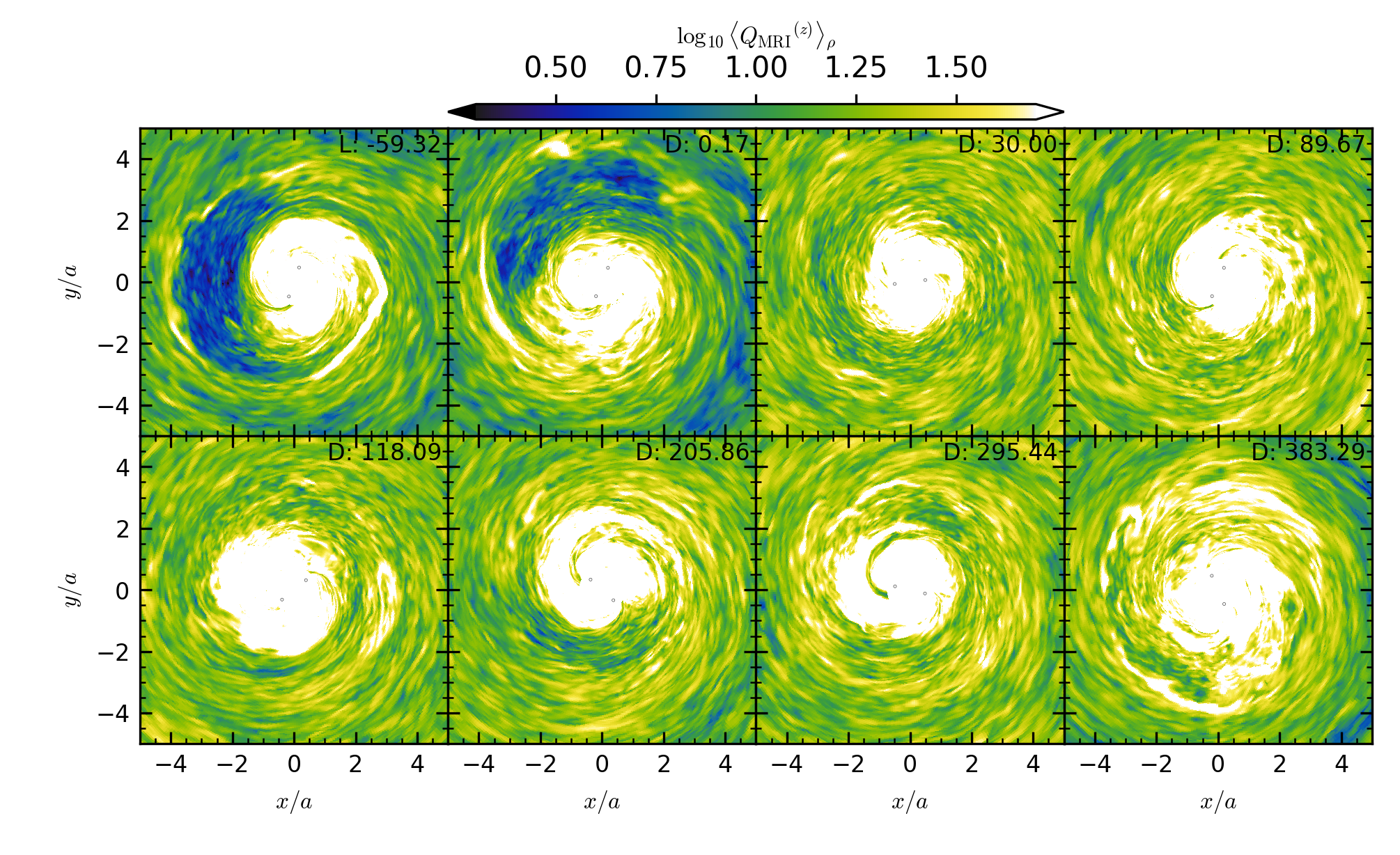}\\[-0.65cm]
    \includegraphics[width=0.7\textwidth]{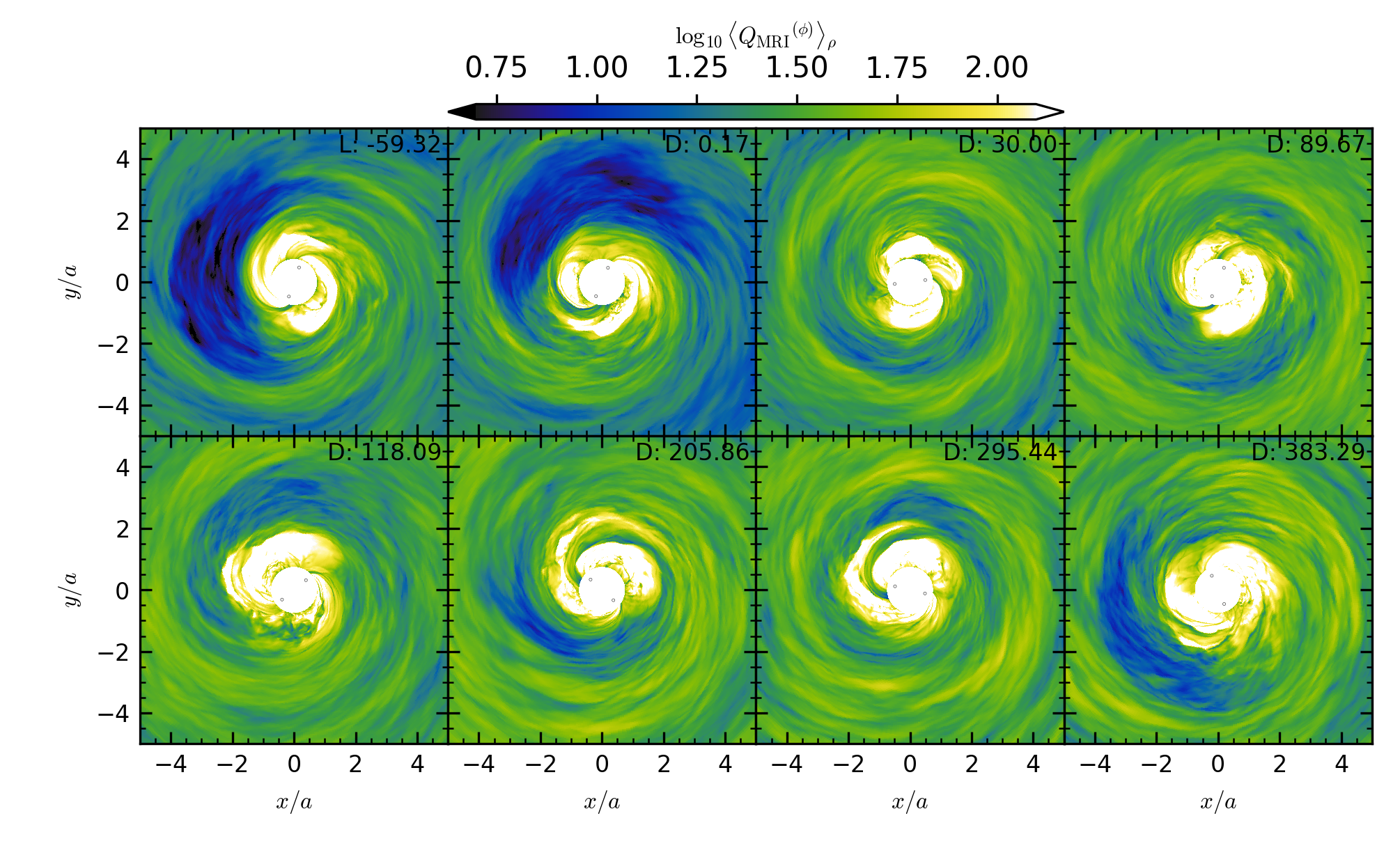}\\[-0.4cm]
    \caption{Snapshots of density-weighted polar average of quality factors at various times for DiffRUN and LegacyRUN. From the top row to the bottom row: vertical quality factor $Q_{\mathrm{MRI}}^{(z)}$, and azimuthal quality factor $Q_{\mathrm{MRI}}^{(\phi)}$. Color is in a logarithmic scale. In the upper right corner of each panel, the letter indicates whether the snapshot is from LegacyRUN(L) or DiffRUN(D), while the number denotes the value of $(t-t_0)/t_{\rm bin}$.}
    \label{fig: QMRI factor}
\end{figure*}

{Maxwell stresses in the CBD are the product of MHD turbulence. Therefore, it is important that the fastest growing mode of the MRI, the primary driver of MHD turbulence, is resolved sufficiently in our simulation.
Our numerical grid is essentially identical to that of Run${_{\rm lrg}}$ \citet{Noble2021}, when normalized by the binary separation. The only difference is that Run${_{\rm lrg}}$ has a $\times 5$ smaller binary separation ($a=20M$) than DiffRUN ($a=100M$), but relativistic corrections are marginal in the disk body, where MHD turbulence is relevant, in both of the simulation.}

{
The resolution quality of a simulation studying MRI-driven nonlinear MHD turbulence can be described in terms of
 the number of cells spanning the wavelength of a fluctuation with wavenumber $k = \Omega_K/V_{A,(i)}$,
}
\begin{equation}
   Q_{\rm MRI}{}^{(i)} \equiv \frac{2\pi |b^i|}{\Delta x^{(i)}~\Omega_{K}(r)~\sqrt{\rho h + 2p_{\rm mag}}}~.
\end{equation}
{Here, the label $i$ refers to either the $\phi$-direction or the $z$-direction,
and $\Delta x^{(i)}$ denotes the cell dimensions in numerical coordinates. $\Omega_{K}(r)$ is the local Keplerian orbital frequency, and $v_{A,i}$ is the Alfven speed corresponding to the magnitude of the magnetic field in the $i$-direction.
\citet{Hawley2011, Hawley2013} recommend that $Q_{\rm MRI}{}^{z}\gtrsim 15$ and $Q_{\rm MRI}{}^{\phi}\gtrsim 20$. 
}

{Our most important results pertain to the mass distribution, so we evaluate global resolution quality in terms of}
a density-weighted average over polar angle:
\begin{equation}
    \langle Q_{\rm MRI}{}^{(i)}\rangle_{\rho}\equiv \frac{\polarint{\rho Q_{\rm MRI}{}^{(i)}}}{\polarint{\rho}}~.
\end{equation}
Fig. \ref{fig: QMRI factor} displays snapshots of these averaged quality factors. %
In LegacyRUN, and at the beginning of DiffRUN, we find that for most of the disk, $\langle Q_{\rm MRI}{}^{z}\rangle_{\rho}$ is $15-40$ and $\langle Q_{\rm MRI}{}^{\phi}\rangle_{\rho}$ is $20-40$.
DiffRUN develops significantly stronger magnetic fields, and therefore better quality factors, than in LegacyRUN: after $30t_{\rm bin}$, $\langle Q_{\rm MRI}{}^{z}\rangle_{\rho}$ is $20-50$ and $\langle Q_{\rm MRI}{}^{\phi}\rangle_{\rho}$ is $30-50$. 
Based on this analysis, we conclude that our simulation sufficiently resolves the MHD turbulence in both LegacyRUN and DiffRUN.}

\clearpage
\bibliography{refs}
\bibliographystyle{aasjournalv7}
\end{document}